\documentclass[10pt,preprint]{aastex}

\begin{document}

\shortauthors{Luhman et al.}
\shorttitle{Census of IC~348}

\title{A Census of the Young Cluster IC~348 \altaffilmark{1}}

\author{K. L. Luhman\altaffilmark{2}, John R. Stauffer\altaffilmark{3}, 
A. A. Muench\altaffilmark{4}, G. H. Rieke\altaffilmark{5}, 
E. A. Lada\altaffilmark{6}, J. Bouvier\altaffilmark{7}, 
and C. J. Lada\altaffilmark{8}}

\altaffiltext{1}{Based on observations obtained at Keck Observatory, Steward 
Observatory, the MMT Observatory, and the Canada-France-Hawaii Telescope.
This publication makes use of data products from the Two Micron All  
Sky Survey, which is a joint project of the University of Massachusetts 
and the Infrared Processing and Analysis Center/California Institute 
of Technology, funded by the National Aeronautics and Space
Administration and the National Science Foundation.}

\altaffiltext{2}{Harvard-Smithsonian Center for Astrophysics, 60 Garden St.,
Cambridge, MA 02138, USA; kluhman@cfa.harvard.edu.}

\altaffiltext{3}{SIRTF Science Center, Caltech MS 314-6, Pasadena, CA 91125,
USA; stauffer@ipac.caltech.edu.}

\altaffiltext{4}{SIRTF Science Center, Caltech MS 220-6, Pasadena, CA 91125,
USA; muench@ipac.caltech.edu.}

\altaffiltext{5}{Steward Observatory, The University of Arizona, Tucson, AZ 
85721; grieke@as.arizona.edu.}

\altaffiltext{6}{Department of Astronomy, The University of Florida, 
Gainesville, FL 32611; lada@astro.ufl.edu.}

\altaffiltext{7}{Laboratoire d'Astrophysique de l'Observatoire de Grenoble, 
BP 53, 38041 Grenoble Cedex 9, France; jbouvier@laog.obs.ujf-grenoble.fr.}

\altaffiltext{8}{Harvard-Smithsonian Center for Astrophysics, 60 Garden St.,
Cambridge, MA 02138, USA; clada@cfa.harvard.edu.}

\begin{abstract}

We present a new census of the stellar and substellar members of the young 
cluster IC~348.
We have obtained images at $I$ and $Z$ for a $42\arcmin\times28\arcmin$ 
field encompassing the cluster and have combined these measurements with
previous optical and near-infrared photometry. 
From spectroscopy of candidate
cluster members appearing in these data, we have identified 122 new members,
15 of which have spectral types of M6.5-M9, corresponding to masses
of $\sim0.08$-0.015~$M_\odot$ by recent evolutionary models.
The latest census for IC~348 now contains a total of 288 members, 23 of which 
are later than M6 and thus are likely to be brown dwarfs.
From an extinction-limited sample of members ($A_V\leq4$) for a 
$16\arcmin\times14\arcmin$ field centered on the cluster, we construct 
an IMF that is unbiased in mass and nearly complete for 
$M/M_{\odot}\geq0.03$ ($\lesssim$M8).
In logarithmic units where the Salpeter slope is 1.35,
the mass function for IC~348 rises from high masses down to a solar 
mass, rises more slowly down to a maximum at 0.1-0.2~$M_\odot$, and then 
declines into the substellar regime. In comparison, the 
similarly-derived IMF for Taurus from Bri\~ceno et al. and Luhman et al.\ rises 
quickly to a peak near 0.8~$M_\odot$ and steadily declines to lower masses.
The distinctive shapes of the IMFs in IC~348 and Taurus are reflected in the 
distributions of spectral types, which peak at M5 and K7, respectively.
These data provide compelling, model-independent evidence for a significant
variation of the IMF with star-forming conditions.

\end{abstract}


\section{Introduction}

The identification of large, unbiased samples of members of star-forming 
regions is important for studying in detail the birth and early evolution
of stars and brown dwarfs. 
Complete membership lists are essential ingredients in the analysis of data 
on circumstellar disks, X-ray emission, multiplicity, rotation, and kinematics.
Most importantly, a thorough spectroscopic census of a young population 
provides a measurement
of the Initial Mass Function (IMF) from massive stars down to brown dwarfs, as
recently demonstrated for Taurus, IC~348, Ophiuchus, and the Trapezium Cluster 
\citep{bri02,luh03,her98,luh98b,luh99,lr99,luh00b,hil97}. 
For measuring the IMF, these spectroscopic surveys complement alternative 
methods such as luminosity function modeling, which have proven successful in 
constraining the IMFs of the more compact and denser clusters 
\citep{mue02,mue03}.

The young cluster IC~348 is well-suited for a membership survey.
The cluster is young (2~Myr), nearby (315~pc), rich ($\sim400$~members),
compact ($D\sim20\arcmin$), and most of its members exhibit relatively low 
extinction ($A_V=0$-4). As a result, a large fraction of the membership can be 
identified down to very low masses through efficient observations of a small
area of sky that can be performed at both optical and infrared (IR) wavelengths.
Previous searches for members of IC~348 have utilized proper motion 
measurements \citep{fre56,sch99}, optical spectroscopy and photometry 
\citep{gin22,har54,str74}, IR photometry \citep{ll95,mue03},
imaging in H$\alpha$ \citep{her54,her98}, 
extensive spectroscopy and photometry at optical and IR wavelengths
\citep{her98,luh98b,luh99}, and narrow-band IR photometry \citep{naj00}
(hereafter NTC00).
The membership information and spectral types from these studies have 
provided an important basis for recent work in IC~348 on multiplicity 
\citep{duc99,lm03}, near-IR and millimeter disk emission 
\citep{hai01,car02,liu03}, variability \citep{her00,rip02}, and
X-ray emission \citep{pre01,pre02}.

In this paper, we present our latest results in a continuing effort to 
identify all of the stellar and substellar members of IC~348. 
We use new data at $I$ and $Z$ and recently published near-IR photometry
to select candidate members of the cluster 
across a larger area and down to lower masses than in previous surveys.
We describe spectroscopy of these candidates and evaluate the membership
status of all objects toward IC~348 that have been observed spectroscopically 
in this work and in previous studies. For this list of known cluster members, 
we estimate extinctions, luminosities, and effective temperatures, and
construct a Hertzsprung-Russell (H-R) diagram. Using the evolutionary
models that provide the best agreement with observational constraints for
young stars \citep{pal99,bar98,cha00} (hereafter BCAH98 and CBAH00), 
we infer individual masses and derive an IMF for the cluster, which is 
compared to the mass functions in other star-forming regions and open clusters.

\section{Observations and Data Analysis}
\label{sec:obs}

\subsection{Optical Photometry}

Images of the IC~348 cluster were obtained with the CFH12K camera on the
Canada-France-Hawaii Telescope on the night of 1999 September 30.
The instrument contained twelve $2048\times4096$ CCDs separated by 
$\sim7\arcsec$ and arranged in a $6\times2$ mosaic. 
With a plate scale of $0\farcs206$~pixel$^{-1}$, the total field of view was
$42\arcmin\times28\arcmin$.  
Images were obtained with exposure times of 1, 30, and 900~s
through the $I$ and $Z$ filters at one pointing centered at
$\alpha=3^{\rm h}44^{\rm m}33\fs2$, $\delta=+32\arcdeg09\arcmin48\arcsec$
(J2000). Data from various targets during the observing run were used to 
construct sky flat field frames. The IC~348 images were bias subtracted 
and flat-fielded. A typical point source in these images exhibited a
FWHM of three pixels, or $0\farcs6$.
Photometry and image coordinates of the sources in these data were 
measured with DAOFIND and PHOT under the IRAF package APPHOT.
For most stars, aperture photometry was extracted with a radius of six pixels.
Smaller apertures were used for closely separated stars. 
The photometry was calibrated in the Cousins $I$ system by combining 
data for standards across a range of colors \citep{lan92} with the 
appropriate aperture and airmass corrections.
The transmission profile for the $Z$ filter is described by \citet{mor03}.
The $Z$ data were calibrated by assuming $I-Z=0$ for A0 standard stars.
This type of calibration is sufficient since the analysis in this study 
relies only on the relative precision of the $Z$ photometry.
To compare our $Z$ photometry to data from another instrument, one would need 
to calibrate the latter in the manner applied to our data and to account for
any differences in the filter and instrument transmission profiles. Saturation 
in the 1~sec exposures occurred near magnitudes of 11-12 in both filters.
The completeness limits of the long exposures were $I\sim22$ and $Z\sim21$.
Typical photometric uncertainties were 0.04~mag at $I=21.25$ and $Z=20.25$ and
0.1~mag at $I=22.25$ and $Z=21.25$. The plate solution was derived from
coordinates measured in the Two-Micron All-Sky Survey (2MASS) Spring 1999 
Release Point Source Catalog for stars that appeared in the optical images 
and were not saturated.

\subsection{Spectroscopy}

The 241 targets of our new spectroscopy are listed in Table~\ref{tab:log},
which are selected in \S~\ref{sec:cand}. The gratings and spectral resolutions 
for these data are also provided in Table~\ref{tab:log}.
During the observations with the Keck low-resolution imaging spectrometer 
(LRIS; \citet{oke95}), we used both the long-slit and multi-slit modes. 
All long-slit spectra were obtained with the slit rotated to the parallactic 
angle. The exposure times ranged 
from 60 to 3600~s. After bias subtraction and flat-fielding,
the spectra were extracted and calibrated in wavelength with arc lamp data.
The spectra were then corrected for the sensitivity functions of the detectors,
which were measured from observations of spectrophotometric standard stars.

In addition to the optical data, we obtained $K$-band spectra of a small
sample of objects using the near-IR long-slit spectrometer FSpec \citep{wil93}. 
These data were collected at the MMT Observatory
on 2000 December 7. Sources were stepped through 
four positions along the slit. At each position, the
integration times ranged between 30 and 120~s. At one grating setting, we
obtained a spectrum extending from 2.0 to 2.4~\micron\ with a two-pixel 
resolution of $R=\lambda/\Delta\lambda=800$. 
We observed a nearby A0~V star (HR~1019), which acted as the telluric standard. 
After dark subtraction and flat-fielding, adjacent images
along the slit were subtracted from each other to remove sky emission.  The
sky-subtracted images were aligned and combined, during which most 
bad pixels and cosmic rays were rejected.  A spectrum was extracted from
this final image, divided by the extracted A0~V standard spectrum, and
wavelength calibrated using OH airglow lines. The intrinsic spectral slope
of the telluric standard was removed with an artificial blackbody spectrum of
$T_{\rm eff}=10000$~K.

Among the several dozen candidate members of IC~348 that were observed 
spectroscopically by \citet{luh99}, only sources with spectral types of M4 or 
later were presented in that study. The 42 stars that were earlier than M4 are 
included in Table~\ref{tab:log}. 
The observations of these stars were conducted on 1998 August 7 and 1998 
December 23 and 26 and are described by \citet{luh99}. 
Table~\ref{tab:log} contains a total of 268 objects, some
of which were observed on more than one date.

\section{New Members of IC~348}
\label{sec:newmem}

\subsection{Selection of Candidate Members}
\label{sec:cand}

\begin{figure}
\epsscale{0.9}
\plotone{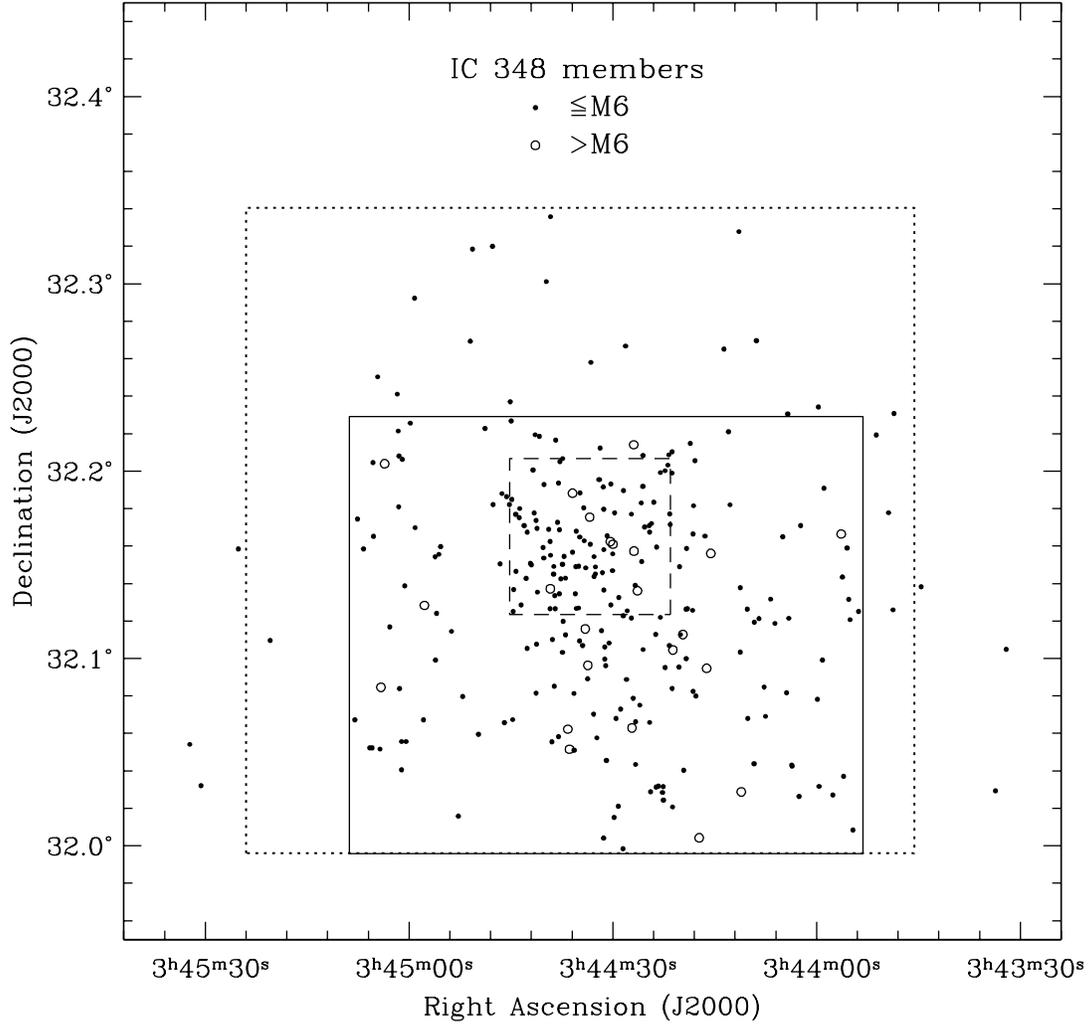}
\caption{
Spatial distribution of the 283 members of the IC~348 cluster that have 
measured spectral types from this work and from previous studies.
The 23 members with spectral types later than M6 are likely to be brown dwarfs. 
Most of the analysis in this work applies to the $16\arcmin\times14\arcmin$ 
field indicated here ({\it solid line}). For reference, the IMF of
\citet{luh98b} was measured for the $5\arcmin\times5\arcmin$ center of the 
cluster ({\it dashed line}), which was closely matched by the region observed 
by NTC00. 
The near-IR imaging and luminosity function modeling of \citet{mue03}
was performed for the $20\farcm5\times20\farcm5$ field ({\it dotted line}).
}
\label{fig:map}
\end{figure}

Previous studies have searched for members of IC~348 through measurements
of proper motions \citep{fre56}, imaging in H$\alpha$ emission 
\citep{her54,her98}, IR luminosity functions \citep{ll95,luh98b,mue03}, 
and optical color-magnitude diagrams \citep{her98,luh99}. 
We extend the latter surveys to greater depth and to a larger area of the 
cluster by including our new $I$ and $Z$ data. We have focused our membership 
survey on the $16\arcmin\times14\arcmin$ region centered at 
$\alpha=3^{\rm h}44^{\rm m}31^{\rm s}$, $\delta=+32\arcdeg06\arcmin45\arcsec$
(J2000), which is indicated in Figure~\ref{fig:map}. The color-magnitude 
diagrams and completeness analysis (\S~\ref{sec:complete}) apply to this area.

To identify the stars toward IC~348 that have both the colors and the 
absolute photometry expected of cluster members, we have constructed 
extinction-corrected diagrams of $I-K_s$ versus $H$ and $I-Z$ versus $H$ in the 
manner described by \citet{luh03}. The near-IR photometry for these 
diagrams is from the references listed in \S~\ref{sec:adopt}.
When the known members of IC~348 are placed on a H-R diagram, most of them 
appear above the model isochrone for an age of 10~Myr from \citet{bar98} 
\citep{luh99}. In the diagrams of $I-K_s$ versus $H$ and $I-Z$ versus $H$ in 
Figure~\ref{fig:col}, we plot the 10~Myr isochrone from 0.015 to 1~$M_{\odot}$ 
by combining the predicted effective temperatures and bolometric luminosities 
\citep{bar98}, a temperature scale that is compatible with the adopted models 
(\S~\ref{sec:teff}), intrinsic colors in $I-K$ \citep{leg92} and $I-Z$ 
(\S~\ref{sec:ext}), dwarf bolometric corrections (see references in 
\citet{luh99}), and a distance modulus of 7.5 \citep{her98}. We have defined 
boundaries below this isochrone to separate candidate members of IC~348 and 
probable field stars, as shown in Figure~\ref{fig:col}.
For spectroscopy, we selected 148 objects that appeared above both of these
boundaries. Although we focused on the $16\arcmin\times14\arcmin$ field in 
Figure~\ref{fig:map}, a small number of the targets are outside of this area. 
We also obtained spectra for 49 objects from previous studies to improve 
their spectral type measurements. 

Young stars that are observed only in scattered light (e.g., edge-on disks)
can appear well below the cluster sequence in a color-magnitude diagram
\citep{bri02}. As a result, such sources are rejected as field stars when
candidate members are selected through color-magnitude diagrams.
However, scattered-light objects can be identified through other indicators
of youth, such as the presence of reflection nebulosity, variability, emission 
in H$\alpha$ and in X-rays, and excess emission at near- and mid-IR wavelengths.
After checking the stars below the boundaries in Figure~\ref{fig:col} for these
signatures, we found that sources 203 and 435 were detected in X-rays by
\citet{pre01} and have near-IR excess emission. The former was also 
detected in the H$\alpha$ survey of \citet{her98}. These two stars were 
included in our spectroscopic sample. 

Many of the above targets were observed with multi-slit spectroscopy.
After designing a slit mask for a set of these objects, we included
additional slitlets as space allowed for stars that were below one or both
of the boundaries in Figure~\ref{fig:col}. The highest priority was given
to stars that appeared to be M-type by the data from NTC00 or
that were just below the boundaries. In this way, useful spectra were obtained 
for 69 objects.

\subsection{Classification of Candidate Members}
\label{sec:class}

In this section, we measure spectral types for the 268 objects toward IC~348
that we observed spectroscopically. 
We then evaluate the membership of each candidate with the spectral types, 
spectral features (H$\alpha$, Na~I, K~I), and photometry.
For determining spectral types and membership, we used low-resolution 
optical spectroscopy from 6000 to 9000~\AA.  At the M spectral types 
that are expected for most of the members of IC~348,
the K~I and Na~I absorption lines vary significantly between dwarfs and
pre-main-sequence stars \citep{mar96,luh98a,luh98b,luh99}. 
Because these features are easily detected in low-resolution spectra,
higher-resolution data for Li were unnecessary for distinguishing young
members from field dwarfs among most of the candidates. Meanwhile, for faint 
M-type sources, both high 
sensitivity and accurate spectral types can be achieved with low spectral 
resolution. For a few objects with earlier spectral types, we included IR
spectroscopy to better constrain the classifications.

\subsubsection{Spectral Types}
\label{sec:sptype}

The low-resolution optical spectra of early-type dwarfs and giants 
($<$K0) exhibit only a few absorption lines (H$\alpha$, Ca~II) and are 
otherwise featureless. As a result, the uncertainties in spectral types 
are largest at these types. In fact, we can classify some objects as only
``early-type" or ``giant". In most of those cases, such a 
classification is sufficiently accurate to indicate that an object is a
background star. However, for G and early K types, we cannot confidently 
distinguish between giants, dwarfs, and pre-main sequence stars
based on the low-resolution optical data. Because absorption in the $K$-band CO 
band heads is much stronger in the former than in the latter two,
we obtained $K$-band spectra of some of the stars with 
these ambiguous classifications and compared the data to those 
of standard dwarfs and giants \citep{lr98} to distinguish 
background giants from dwarfs and pre-main-sequence stars.
In addition, we measured the strength of Li absorption at 6707~\AA\ for one 
these stars through moderate-resolution spectroscopy.
There remains one star of this kind that lacks IR spectra or Li data and 
thus has an uncertain classification (\S~\ref{sec:append1a}).

\begin{figure}
\epsscale{0.9}
\plotone{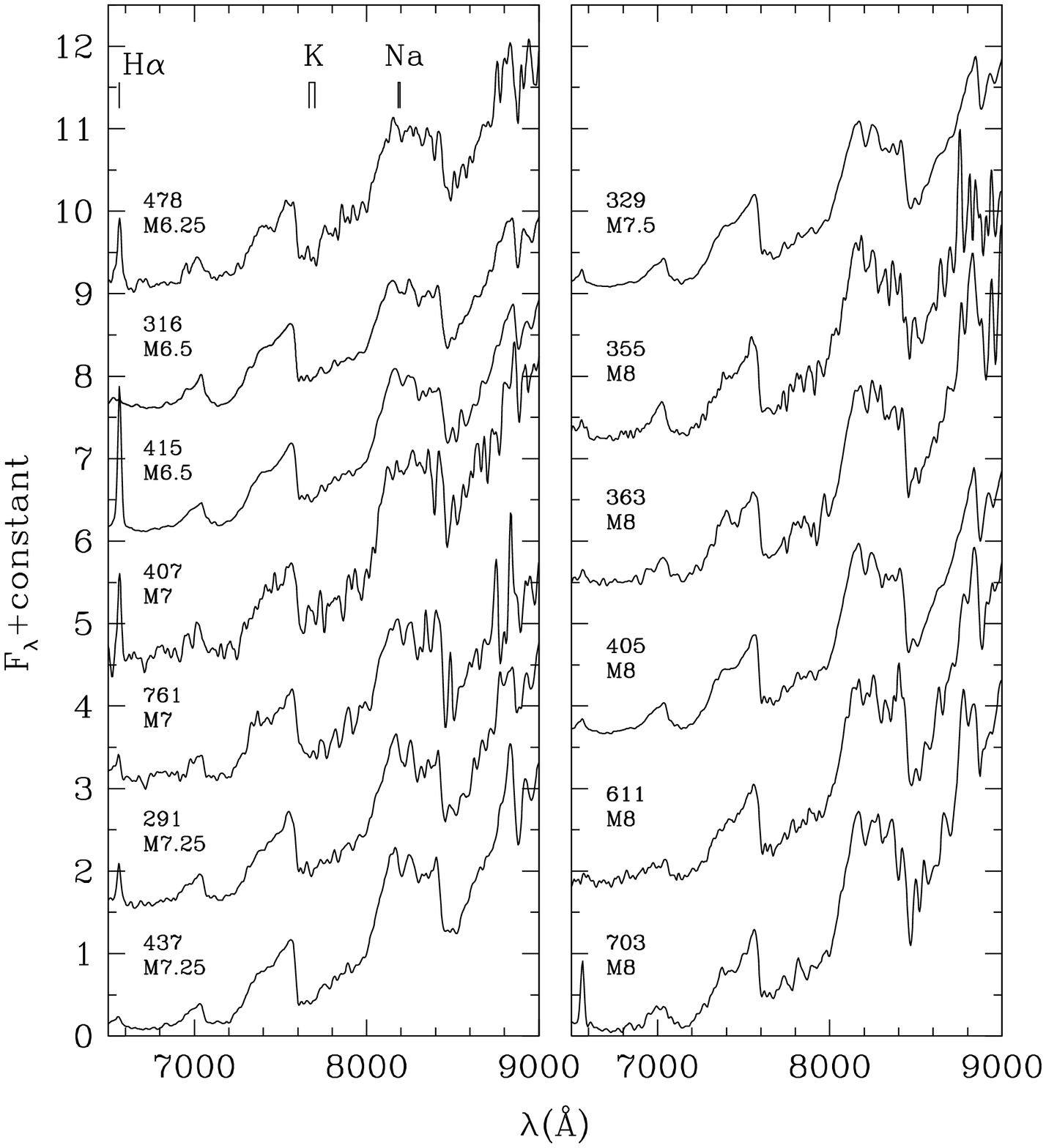}
\caption{
Low-resolution spectra of the 13 known members of the IC~348 cluster with
spectral types of M6.5 to M8 (\citet{luh98b}; \citet{luh99}; this work), which
are below the hydrogen burning mass limit by the H-R diagram and evolutionary 
models in Figure~\ref{fig:hr2}. These spectra exhibit reddenings of $A_V=0$-4.
All data are smoothed to a resolution of 25~\AA\ and normalized at 7500~\AA.
}
\label{fig:spec1}
\end{figure}

\begin{figure}
\epsscale{0.9}
\plotone{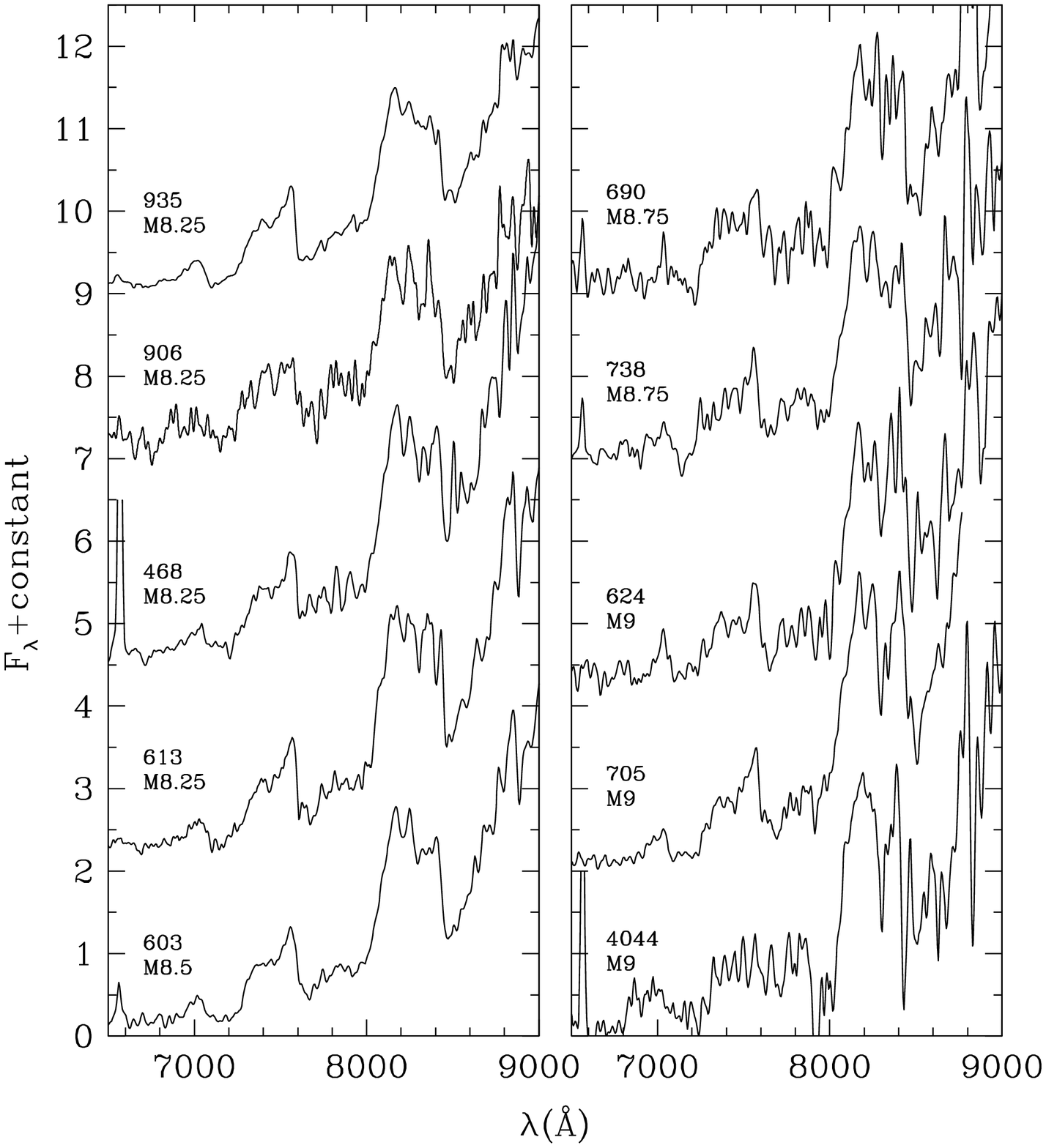}
\caption{
Low-resolution spectra of the 10 known members of the IC~348 cluster with
spectral types of M8.25 to M9 (\citet{luh98b}; \citet{luh99}; this work), which
are below the hydrogen burning mass limit by the H-R diagram and evolutionary 
models in Figure~\ref{fig:hr2}. These spectra exhibit reddenings of $A_V=0$-2.
All data are smoothed to a resolution of 25~\AA\ and normalized at 7500~\AA.
}
\label{fig:spec2}
\end{figure}

For K and M-type field dwarfs and for K0-M5 members of IC~348, we measured 
spectral types by comparison to the spectra of standard dwarfs from 
\citet{as95}, \citet{kir91}, \citet{hen94}, and \citet{kir97}.
Spectral types for members later than M5 were derived with the averages of 
dwarf and giant spectra as described by \citet{luh99}. The same system has 
been used in classifying young late-type objects in Taurus \citep{bri02,luh03}, 
MBM~12A \citep{luh01}, Ophiuchus \citep{luh97}, and IC~348 \citep{luh98b,luh99}.
Spectra of the 23 known members of IC~348 with spectral types later
than M6 are presented in Figs.~\ref{fig:spec1} and \ref{fig:spec2}.
The typical uncertainties in the optical spectral types from \citet{luh98b},
\citet{luh99}, and this work are $\pm0.5$ and $\pm0.25$ for
K and M types, respectively, unless noted otherwise.

\subsubsection{Membership}
\label{sec:mem}

Stars projected against IC~348 can be background stars, foreground stars, or 
young members of the cluster.  We now evaluate the membership status of 
all objects toward IC~348 that have been observed spectroscopically in this 
work and in previous studies.

Members of IC~348 can be identified by several indicators.
Nine of the brightest stars toward IC~348 have relative proper motions that 
imply membership in the cluster \citep{fre56}. Meanwhile, the deeper proper 
motion measurements of \citet{sch99} cannot reliably distinguish members
from background stars. Because IC~348 is a star-forming cluster, objects that
show evidence of youth are likely to be members of the cluster. Signatures of
youth include H$\alpha$ emission above the levels observed for active 
field dwarfs ($W_{\lambda}\gtrsim15$~\AA\ for M types), Br$\gamma$ 
emission, IR excess emission in photometry or spectra, and 
spectral features implying low gravity. Examples of the latter are K~I, Na~I, 
and the overall shape of the red optical spectra, which differ noticeably
between dwarfs and pre-main-sequence objects at spectral types later than M2.  
Finally, for a given star, the presence of significant reddening in 
the spectrum or colors ($A_V>1$) and a position above the main sequence for the 
distance of IC~348 indicate that it cannot be in the foreground or the 
background of the cluster, respectively, and therefore must be a member of the 
cluster. 

Field dwarfs in the foreground of IC~348 have large proper motions relative
to cluster members or background stars, and thus are easily identified 
in proper motion surveys \citep{bla52,sch99}. 
Foreground dwarfs are also characterized by little or no reddening,
strong absorption in the optical Na~I and K~I transitions 
at M types, and a lack of Li absorption or any other indicator of youth.

Field stars in the background of IC~348 consist of dwarfs and giants.
At brighter levels, most background stars are giants and early-type dwarfs.
However, because of the depth of our spectroscopy, we detected a large number 
of background field M dwarfs as well. 
Background dwarfs fall below the main sequence
if placed on the H-R diagram for the distance of the cluster. 
These M dwarfs also can be identified by 
their strong absorption in Na~I and K~I, just as for foreground dwarfs.
For some of the M-type stars falling below the main sequence, we
cannot use these features to confirm their nature as field dwarfs because of
insufficient signal-to-noise in the spectra. As discussed in \S~\ref{sec:cand},
young stars that are seen in scattered light can appear low on the H-R diagram, 
even below the main sequence. As a result, some of the objects that we identify 
as background M dwarfs could be scattered-light members of IC~348. However,
these stars show none of the signatures of youth mentioned 
at the end of \S~\ref{sec:cand}.

We list the spectral types from this work and from previous studies for objects 
that we classify as members of IC~348, as foreground stars, and as background
stars in Tables~\ref{tab:mem}, \ref{tab:fore}, and \ref{tab:back},
respectively. The evidence for the assigned membership status of each
member and foreground star is provided. The background
stars were identified by either a position below the main sequence and a lack of
youth indicators or by a classification as a giant. 
Stars that have uncertain membership status 
by available data are found in Table~\ref{tab:unc}. In Table~\ref{tab:names},
our source identifications are listed with those from \citet{her98},
NTC00, \citet{pre01}, and the 2MASS Point Source Catalog. A few mistakes in the 
cross-identifications from NTC00 and \citet{pre01} have been corrected 
here. In our new spectroscopic sample of 268 objects, there are 47 members that
have previously published spectra, 122 newly confirmed members (15 later than
M6), 97 field stars, and two stars with uncertain membership status. 
Additional comments on the spectral and membership classifications of 
individual sources are in \S~\ref{sec:append1a}.

The membership list in Table~\ref{tab:mem} contains 288 entries.
Three stars are close secondaries that lack resolved spectroscopy.
Source 51 exhibits a featureless IR spectrum that is indicative of a young
star with a class~I spectral energy distribution. The spectrum of object 46 
confirms its youth and membership, but did not have sufficient signal-to-noise 
for an accurate spectral classification. The remaining 283 members have 
measured spectral types. The positions of these objects are plotted on the map 
in Figure~\ref{fig:map}.

\subsection{Completeness of Census}
\label{sec:complete}

In the following discussion, we assess the completeness of the new
census of members within the $16\arcmin\times14\arcmin$ field toward 
IC~348 that is shown in Figure~\ref{fig:map}.

\begin{figure}
\plotone{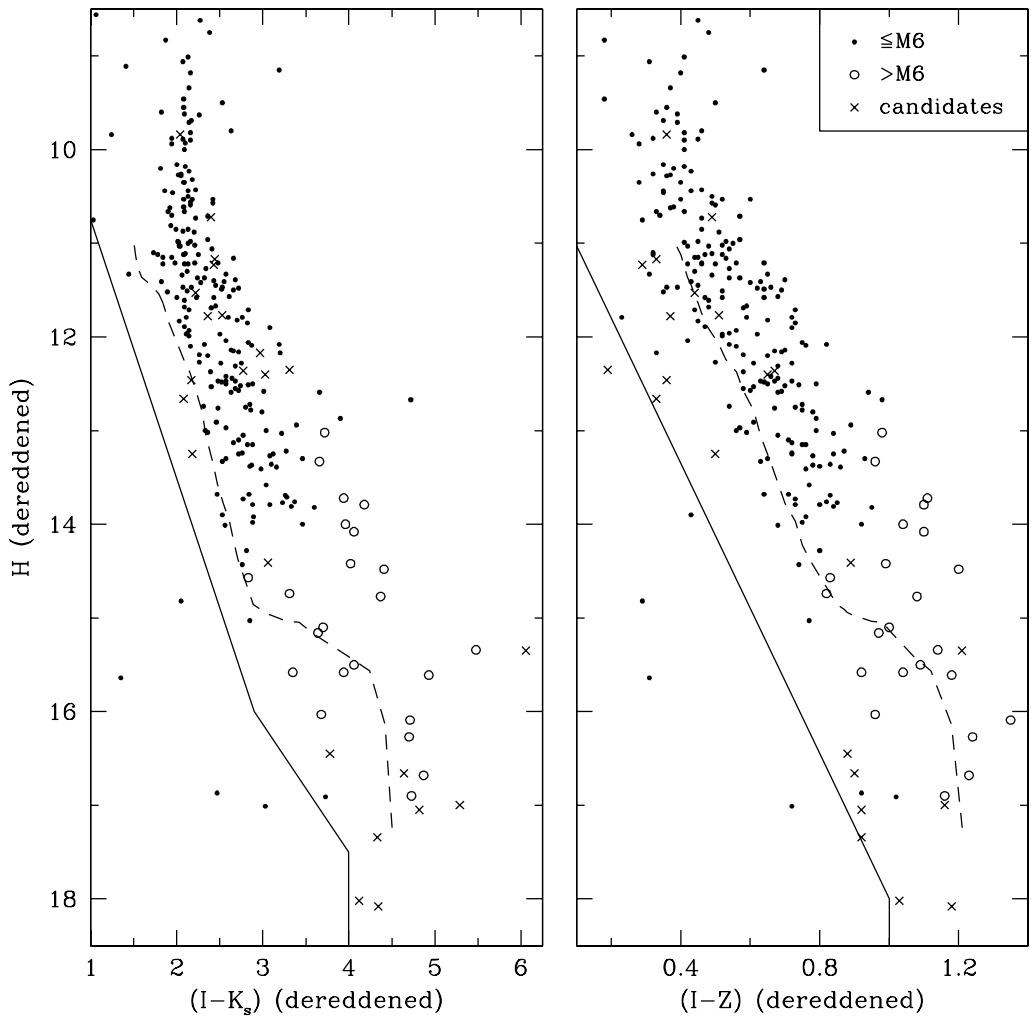}
\caption{
Extinction-corrected color-magnitude diagrams for the 
$16\arcmin\times14\arcmin$ field in the IC~348 cluster in Figure~\ref{fig:map}.
We show the cluster members at $\leq$M6 and $>$M6 ({\it points and circles})
that have been identified through spectroscopy in this work and in previous 
studies.
We have omitted the foreground and background stars that appear in spectroscopic
and proper motion measurements (Tables~\ref{tab:fore} and \ref{tab:back})
as well as objects that are likely to be field stars by their location 
below either of the solid boundaries in this diagram.
Sources that lack spectroscopic measurements and that are above both of the 
boundaries are candidate members of IC~348 ({\it crosses}).
One star that is below the boundary in the right diagram shows other evidence 
of youth and thus is marked as a candidate (\S~\ref{sec:cand}).
We omit stars that are too faint to be reliably identified as either 
field stars or candidate members ($I>22$).
The dashed line is the 10~Myr isochrone (1-0.015~$M_{\odot}$) from the 
evolutionary models of \citet{bar98}.
}
\label{fig:col}
\end{figure}

We first examine our photometry for candidate cluster members that have 
not been observed spectroscopically.
In Figure~\ref{fig:col}, we indicate the confirmed members from the 
previous section, while omitting the known field stars. 
Next, we use this diagram to identify candidate members among the stars 
that lack spectroscopy in the $16\arcmin\times14\arcmin$ field.
At $I>22$, we cannot use the boundaries in Figure~\ref{fig:col} to 
reliably separate field stars and candidate members because the photometric 
uncertainties are too large. Therefore, we apply this diagram only to stars
with $I\leq22$.
Objects that are above this limit and that fall below either of the boundaries 
in Figure~\ref{fig:col} are likely to be field stars, and thus are not shown.
Similarly, we also omit stars that appear below the boundary in the 
diagram of $R-I$ versus $I$ from \citet{luh99}.
The one exception is 2MASS~03444520+3201197, which falls below the boundary in
$I-Z$ versus $H$, but is surrounded by reflection nebulosity in our $I$ and
$Z$ images, and thus is probably a cluster member observed in scattered light. 
Sources that are earlier than M2 by the narrow-band photometry of NTC00
and that have $K=14$-16 are likely field stars as well 
(see Figure~\ref{fig:nic}).
After rejecting these field stars, in Figure~\ref{fig:col} there 
remain 23 candidate members in the $16\arcmin\times14\arcmin$ field
that have not been observed with spectroscopy. In comparison, 261 cluster 
members have been confirmed in this region in this work and previous studies
(\S~\ref{sec:mem}).

\begin{figure}
\plotone{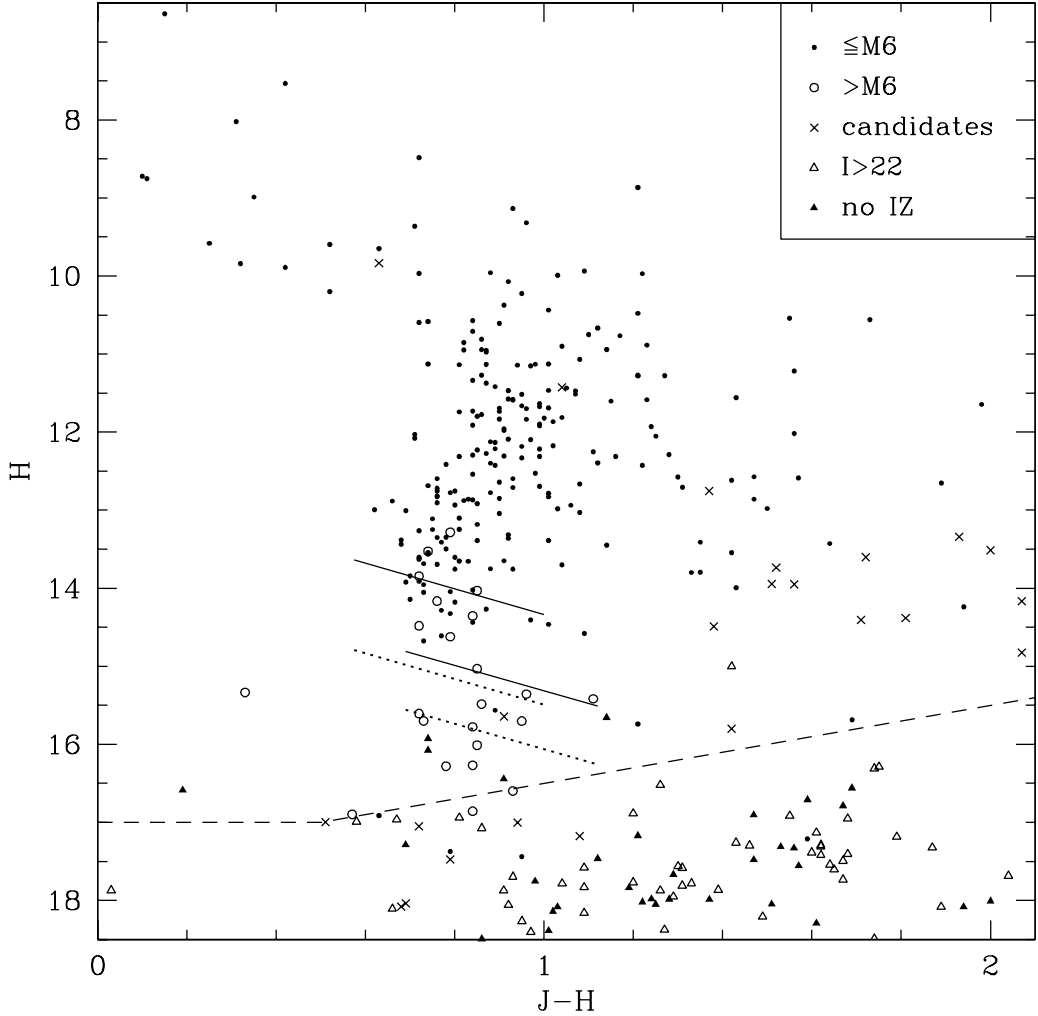}
\caption{
$J-H$ versus $H$ for the $16\arcmin\times14\arcmin$ field in the IC~348 cluster
in Figure~\ref{fig:map}.
The symbols are the same as in Figure~\ref{fig:col}, with the addition of 
stars with uncertain optical photometry ($I>22$, {\it open triangles})
and sources detected only in these IR data ({\it solid triangles}).
The reddening vectors from $A_V=0$-4 are plotted for 0.08 ($\sim$M6.5) and 
0.03~$M_\odot$ ($\sim$M8) for ages of 3~Myr ({\it upper and lower solid lines})
and 10~Myr ({\it upper and lower dotted lines}) (CBAH00). 
Most of these $J$ and $H$ measurements are from 2MASS and \citet{mue03}.
The dashed line represents the completeness limits for the data from 
\citet{mue03}.
}
\label{fig:jh}
\end{figure}

Because the mass and reddening vectors are roughly perpendicular in a
near-IR color-magnitude diagram and because cool and reddened sources are
most easily detected at near-IR wavelengths, we use the diagram of $J-H$ 
versus $H$ in Figure~\ref{fig:jh} to evaluate the completeness of the 
current census in terms of mass and extinction. The photometry in 
Figure~\ref{fig:jh} is compiled from references in \S~\ref{sec:adopt}. 
The deepest of those data sets is from \citet{mue03}, which 
encompassed the entire $16\arcmin\times14\arcmin$ field. 
The completeness limits of the photometry from \citet{mue03} are 
taken to be the magnitudes at which the logarithm of the number of sources as a 
function of magnitude departs from a linear slope and begins to turn over 
($J\sim17.5$, $H\sim17.0$, $K_s\sim16.5$). In Figure~\ref{fig:jh}, we have 
plotted the reddening vectors from $A_V=0$-4 for cluster members with masses 
of 0.03 and 0.08~$M_{\odot}$ and ages of 3 and 10~Myr,
where the latter is the maximum age implied by the H-R diagram
of the cluster (\S~\ref{sec:hr}).
These vectors are derived in the manner described by \citet{bri02}
with the evolutionary models of CBAH00 for the distance of IC~348.
As demonstrated in Figure~\ref{fig:jh}, the IR photometric data should be 
complete for members of IC~348 with $M/M_{\odot}\geq0.03$ and $A_V\leq4$,
with the exception of close companions.

In Figure~\ref{fig:jh}, we omit known and likely field stars in the 
same manner as in Figure~\ref{fig:col}. The remaining IR sources consist of
confirmed members and objects that lack spectroscopic data. The latter are 
divided into candidate members from Figure~\ref{fig:col}, stars with uncertain 
optical photometry ($I>22$), and sources detected only in the IR data. 
In Figure~\ref{fig:jh}, there are only three objects that lack spectral
classifications and that 
would have $M/M_{\odot}>0.03$ and $A_V\leq4$ if they were members ($H<16$, 
$J-H<1.2$), all of which are candidate members by Figure~\ref{fig:col}.
Because of the small number of these candidates, the current sample of 
confirmed members is nearly complete for these extinction and mass limits.
Therefore, in \S~\ref{sec:imf}, we will derive the IMF from a sample of
known members defined by $A_V\leq4$ and we will refer to 0.03~$M_{\odot}$ 
as the completeness limit.
Just as this survey is incomplete for low-mass members with high
extinction ($A_V>4$), it is not sensitive to objects that are seen
in scattered light (e.g., edge-on disks), particularly those at low-masses.

\subsection{Implications for the Survey of NTC00}
\label{sec:najita}

\begin{figure}
\plotone{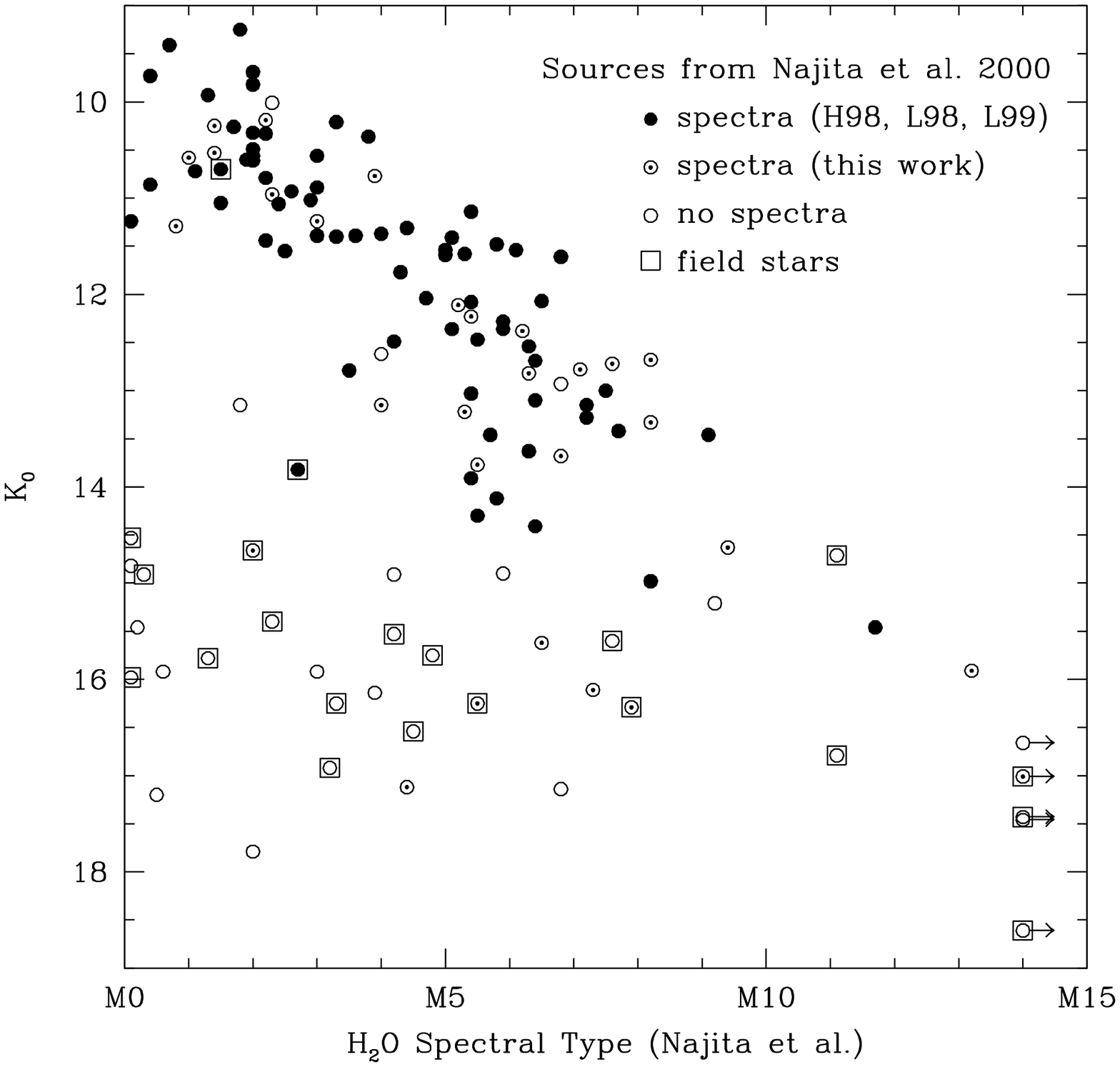}
\caption{
Spectral types estimated from $HST$ NICMOS narrow-band photometry of H$_2$O 
absorption versus dereddened $K$-band data for the $5\arcmin\times5\arcmin$
center of IC~348, as reported by NTC00. 
We indicate sources that have been observed spectroscopically by 
\citet{her98}, \citet{luh98b}, and \citet{luh99} ({\it large points}) 
and in this work ({\it circled points}), sources that lack spectra
({\it circles}), and objects that are identified as foreground or
background field stars by spectroscopy or the color-magnitude diagrams in 
Figure~\ref{fig:col} and in \citet{lm03} ({\it squares}).
}
\label{fig:nic}
\end{figure}

Using the spectral type and membership data compiled in \S~\ref{sec:class},
we evaluate the survey for low-mass members of IC~348 by NTC00.
In that study, the $5\arcmin\times5\arcmin$ center of the cluster was imaged 
with NICMOS on $HST$ in narrow-band filters that sampled the near-IR steam 
absorption bands. 
They estimated spectral types from those data and constructed an empirical
H-R diagram, which is reproduced in Figure~\ref{fig:nic}. From that diagram,
they identified 20-30 brown dwarfs candidates ($>$M6).

We first discuss the membership status of the sources detected by NTC00.
We place their observations in the context of
previous surveys by indicating in Figure~\ref{fig:nic} the sources that had 
already been observed spectroscopically by \citet{her98}, \citet{luh99},
and \citet{luh98b}. We mark the objects that have spectra for 
the first time from this work, two of which, 603 and 624, had been previously 
reported as candidate low-mass members \citep{luh00b}.
Next, we indicate the NICMOS sources that are field stars by the spectroscopic 
analysis of the previous section or by the positions in the color-magnitude 
diagrams in Figure~\ref{fig:col} and in \citet{lm03}. Several objects that 
appear to have spectral types of late M by the NICMOS data -- and thus were 
taken as probable low-mass members by NTC00 -- are field stars. 
Sources 609, 618, and 1426 have NICMOS spectral types of 
M7.9$\pm3.9$, M5.5$\pm2.9$, and $>$M14, respectively, but are background
stars by both spectroscopy and the color-magnitude diagrams.
While NICMOS objects 021-05, 022-03, 052-05, 071-02, and 075-01 were 
classified as M11.1$\pm5.2$, $>$M14, $>$M14, M7.6$\pm6.7$, and M11.1$\pm6.6$ 
by NTC00, they are probably field stars by their positions 
in an optical color-magnitude diagram constructed from $HST$ WFPC2 data 
by \citet{lm03}.
Finally, we examine the objects that are not labeled as field stars and 
that lack spectra. 
Three NICMOS sources near the cluster sequence at $K<14$ have not been observed
spectroscopically. Two of them are the close companions 60B and 78B. 
One other object, 073-01 from NTC00,
is a candidate member by our optical color-magnitude diagrams. 
At $K>14$, the remaining candidate members by the NICMOS data are 
014-06, 024-02, 072-01, 082-02, and 104-01 which have NICMOS 
types of M6.8$\pm12.1$, M9.2$\pm4.8$, $>$M14, $>$M14, and M5.9$\pm3.9$. 
Given the large uncertainties in these 
classifications and the fact that these stars are not detected in our 
optical images, they could be either reddened background stars ($A_V\gtrsim3$) 
or cool sources that are intrinsically red (field dwarfs or members).

\begin{figure}
\plotone{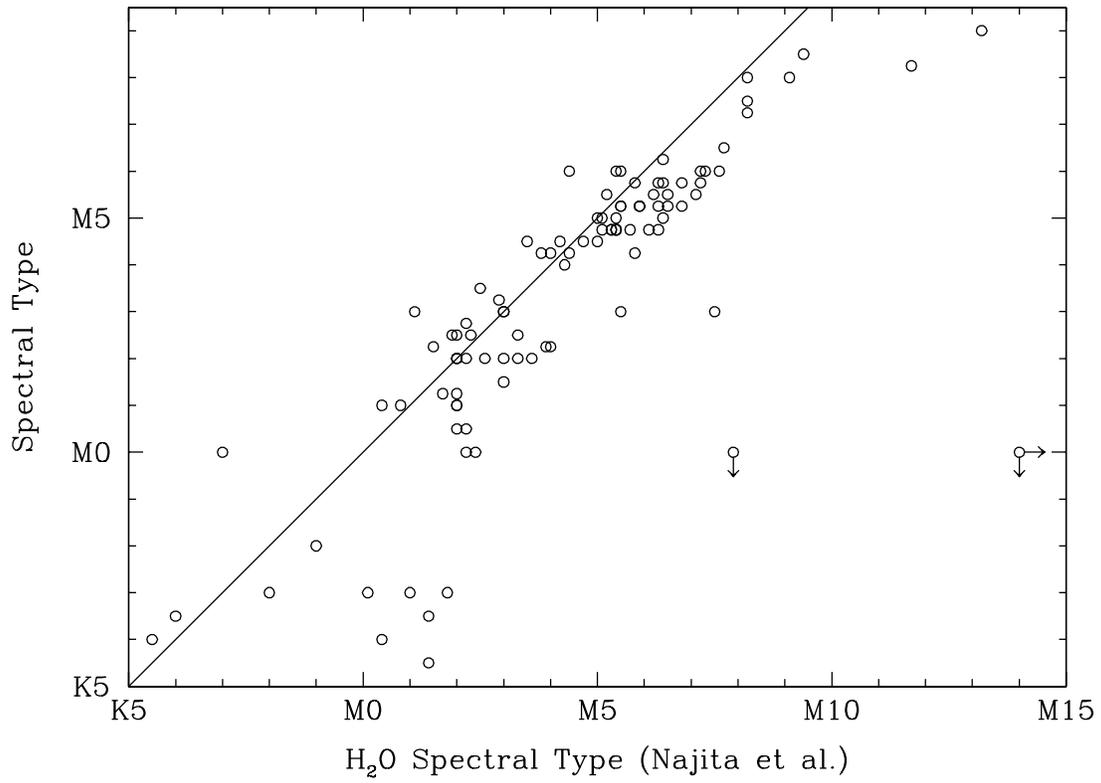}
\caption{
Spectral types estimated from $HST$ NICMOS narrow-band photometry of H$_2$O 
absorption by NTC00 versus spectral types measured from optical 
spectra by \citet{her98}, \citet{luh98b}, and \citet{luh99} and in this work. 
The typical uncertainties in the spectral types on the vertical axis 
are $\pm0.25$-0.5~subclass.
}
\label{fig:comp}
\end{figure}

We now investigate the accuracy and precision of the spectral types that 
were estimated from the NICMOS narrow-band photometry by NTC00.
In Figure~\ref{fig:comp}, we plot spectral types from NTC00
and spectral types measured from ground-based spectroscopy for all NICMOS 
sources for which the latter data are available. Most of the ground-based 
spectral types are derived from optical data. The small number of 
IR types included here were measured with optically-classified standards. 
Therefore, the ground-based classifications are assured to be accurate, and 
in fact effectively define the spectral types for young objects at late M types.
As for the precision of the optical spectral types, the measurement
errors are $\pm1$ subclass or less for most of the objects in question. 
The accuracy of the NICMOS spectral types is illustrated in 
Figure~\ref{fig:comp}, where the NICMOS classifications are
systematically later than those measured from the optical data by an 
average of $\sim1$-2~subclasses. 
The spectral type calibration of the NICMOS measurements was derived
from both optically-classified members of IC~348 and field dwarf standards.
The offset in classifications in Figure~\ref{fig:comp} is probably
due to differences in the steam band strengths between dwarfs and young objects
at a given spectral type, which is a behavior that has been noticed
previously \citep{lr99,luc01,all01}.
The precision of the NICMOS spectral types is quantified by the measurement 
errors reported by NTC00, which are largest for early types and 
for faint objects; the NICMOS types at $<$M2 are uncertain because little 
steam absorption is present at higher temperatures, as pointed out by NTC00, 
while the increase in errors at faint levels is
a reflection of photometric uncertainties.

As demonstrated in the novel study of NTC00, measurements of near-IR steam 
absorption through narrow-band photometry can be used to efficiently identify
cool objects that may be low-mass members of a dense cluster. 
However, to arrive at an accurate calibration of spectral types,
optically-classified young M-type objects rather than M dwarfs should be
used as the spectral type standards. 
As narrow-band photometry does not distinguish between cool field stars and
cool members, membership of candidates should be confirmed through 
other means (e.g., spectroscopy, proper motions). In addition, useful 
estimates of spectral types can be derived only well above the detection 
limit where the photometry is accurate.

\section{The IC~348 Stellar Population}
\label{sec:pop}

In this section, we begin by tabulating photometry and spectral types for all 
known members of the IC~348 cluster and estimating their
extinctions, effective temperatures, and bolometric luminosities 
(\S~\ref{sec:adopt}, \S~\ref{sec:ext}, \S~\ref{sec:teff}). 
The latter three parameters are estimated for the 283 members
of IC~348 in Table~\ref{tab:mem} that have resolved spectral types.
We place these sources on the H-R diagram and interpret their positions 
with the most suitable set of theoretical evolutionary models (\S~\ref{sec:hr}).
Because the current census of members is unbiased in mass at
$M/M_{\odot}\geq0.03$ and $A_V\leq4$ for the $16\arcmin\times14\arcmin$
region shown in Figure~\ref{fig:map}, we will restrict the IMF sample
to include only members within this extinction limit and in this area. 
The resulting IMF is compared to data from other stellar populations such as
the Taurus star-forming region (\S~\ref{sec:imf}).
Finally, we discuss the implications of our new data for the X-ray observations 
of \citet{pre01} (\S~\ref{sec:xray}).

\subsection{Adopted Data}
\label{sec:adopt}

The latest membership list for IC~348 was compiled in \S~\ref{sec:mem} and 
is given in Table~\ref{tab:mem}.  High-resolution imaging has resolved some
of these sources into close binaries. To facilitate comparisons of the 
populations in IC~348 and other young regions, we treat a binary system in 
IC~348 with a separation less than $1\arcsec$ as one object for the remainder 
of this section.
In order of preference, we adopt the $R-I$ colors from \citet{luh99} and
\citet{her98}, the $I$-band measurements from this work, \citet{luh99},
and \citet{her98}, and the near-IR data from \citet{mue03},
\citet{luh98b}, the 2MASS Point Source Catalog, and \citet{ll95}.
For reasons given in \citet{luh99}, we do not use the
$R-I$ colors of \citet{her98} for $R-I\geq1.5$. 
The $J$, $H$, and $K$ measurements of \citet{luh98b} are an average of
0.09, 0.19, and 0.16~mag fainter than the 2MASS photometry for sources in common
between the two data sets. We subtract these offsets from the data of 
\citet{luh98b} for this study. 
The data from \citet{mue03} are used only when $J>10$, $H>10.5$, and
$K_s>11$, which are below the saturation limits of that survey. 
We adopt the coordinates measured from our $I$-band images and otherwise use
the values from 2MASS. Coordinates are not available for source 91 
from either of these sets of data. For this star, we measure the average offset
between the coordinates of stars in the image containing 91 from NTC00
and in our $I$-band images. We use this offset and 
the coordinates reported for 91 by NTC00 to arrive at
coordinates that are on the reference frame of our $I$-band images.
These adopted measurements, the $I-Z$ colors from this 
work, and the available spectral types are presented in Table~\ref{tab:mem}.

\subsection{Extinctions}
\label{sec:ext}

In the following analysis, standard dwarf colors are taken from the compilation 
of \citet{kh95} for types earlier than M0 and from the young
disk populations described by \citet{leg92} for types of M0 and later.
The IR colors from \citet{kh95} are transformed from the 
Johnson-Glass photometric system to the CIT system \citep{bb88}.
Near-IR colors in the 2MASS and CIT photometric systems agree at a level of 
$<0.1$~mag \citep{car01}. 
From the distributions of $E(R-I)$ and $E(I-Z)$ versus $E(J-H)$ produced
later in this section, we inferred 
$E(I-Z)=0.77$~$E(J-H)$ and $E(R-I)=1.7$~$E(J-H)$.
When these relations were combined with the extinction law of \citet{rl85}, we 
arrived at $E(I-Z)=0.082$~$A_V=0.29$~$A_J$ and $E(R-I)=0.18$~$A_V=0.64$~$A_J$.


The amount of extinction towards a young star can be estimated from the 
reddening of its broad-band colors. To ensure that the color excesses 
reflect only the effect of extinction, contamination from short and long 
wavelength excess emission is minimized by selecting colors between the $R$ and 
$J$ bands. Therefore, our extinction estimates are based on $R-I$ and $I-Z$. 
To measure extinctions from the observed colors, we need the intrinsic
values of $R-I$ and $I-Z$ as a function of spectral type, which were estimated
in the following manner.
Although $J-H$ is more susceptible to contamination from emission from 
circumstellar material than $R-I$ and $I-Z$, only a small minority of the
members of IC~348 exhibit such excess emission in $J-H$. In addition, 
the intrinsic photospheric $J-H$ colors of members of IC~348 were shown 
to be dwarf-like by \citet{luh99}. As a result, by assuming dwarf intrinsic
colors for the sources with no near-IR excess emission, we were able to 
calculate reliable extinctions for a large sample of members. We then 
combined these extinctions with the observed $R-I$ and $I-Z$ for IC~348 members
to arrive at average intrinsic colors as a function of spectral type. 
At M2 through M5, these estimates of $R-I$ were bluer than the values for 
dwarfs, which is a departure toward giant-like colors. We found a similar 
result in a previous study of a smaller sample of sources in IC~348
\citep{luh99}.
For other spectral types, the $R-I$ colors that we derived were consistent 
with dwarf colors, and so we adopted the latter at these types.
The intrinsic $R-I$ and $I-Z$ at each spectral type from K0 
through M9 are listed in Table~\ref{tab:iz}. 
Because of the paucity of members earlier than K0, we did not 
estimate the average intrinsic $I-Z$ at those types.
Since $R-I$ data are not available for many of the members later 
than M6, we also measured reddenings from the optical spectra during the 
process of spectral classification of these sources.
The final extinctions are averages 
of the values implied by $R-I$, $I-Z$ ($\geq$K0), and the optical spectra 
($>$M6).  None of these three measurements are available for 23 sources, which 
are either saturated or below the detection limit in the optical data or 
are earlier than K0. 
For these stars, we measured extinctions from $E(J-H)$ assuming dwarf-like
intrinsic colors. One exception is source
13, which exhibits strong IR excess emission. The extinction for this object
was estimated by dereddening the $J-H$ and $H-K_s$ colors to the locus
observed for classical T~Tauri stars \citep{mey97}. 



\subsection{Effective Temperatures and Bolometric Luminosities}
\label{sec:teff}

\begin{figure}
\plotone{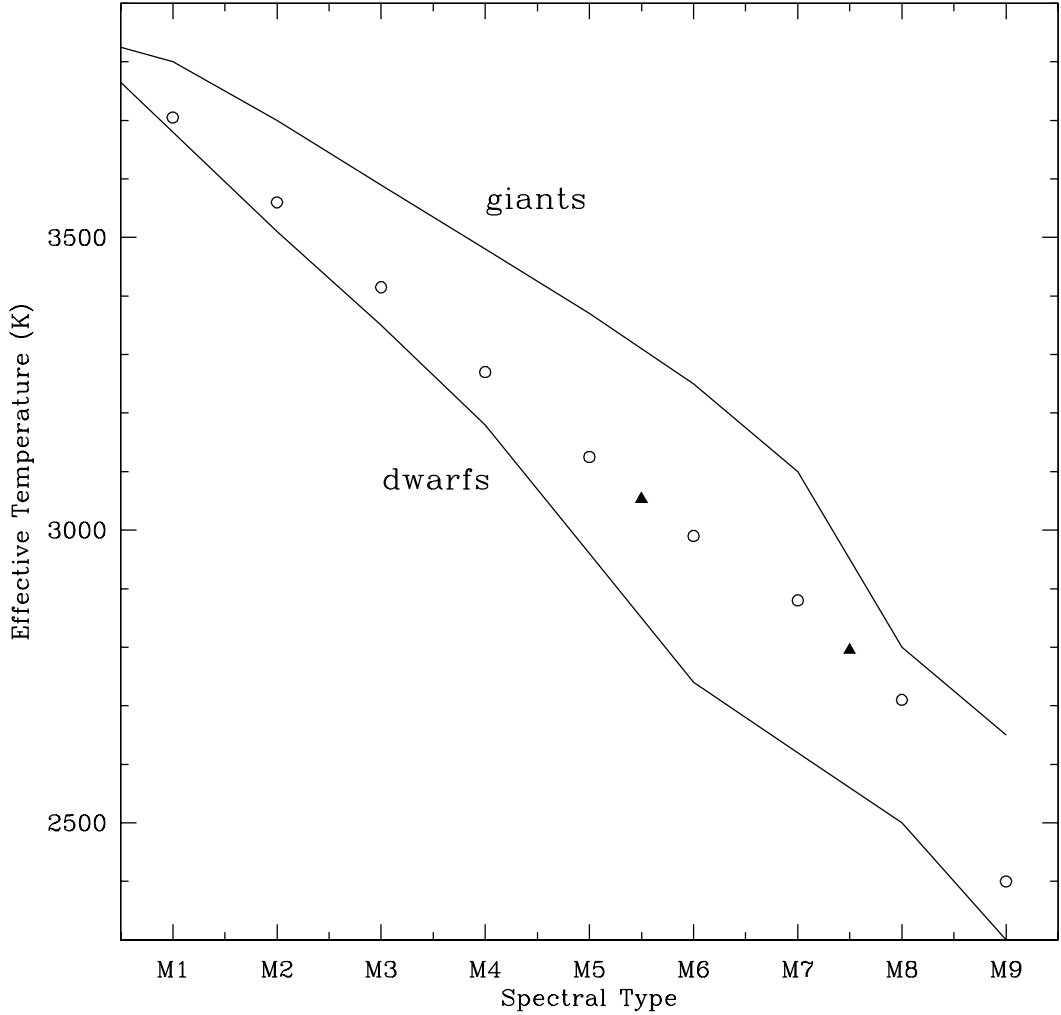}
\caption{
For GG~Tau~Ba and Bb (M5.5, M7.5) to have the same age as GG~Tau~Aa and Ab 
(K7, M0.5) on the model isochrones of BCAH98 and CBAH00,
the first two objects must have the temperatures indicated
by the solid triangles. We have constructed a temperature scale 
({\it circles}; Table~\ref{tab:scale}) that coincides with these 
two points and thus produces coevality for GG~Tau.
At $>$M7.5, the scale has been designed so that the M8-M9 members of Taurus 
and IC~348 have model ages that are similar to those of the earlier members.
Temperature scales for dwarfs and giants are shown for comparison
({\it solid lines}). The dwarf scale compiled by \citet{luh99} has been
adjusted by $-50$~K at M5 and $-100$~K at M6-M9 to be consistent with the
latest temperature estimates for young disk dwarfs 
\citep{leg00,leg01,bur02}. The references for the giant scale are provided
in \citet{luh99}.
}
\label{fig:scale}
\end{figure}

Spectral types of M0 and earlier are converted to effective temperatures with
the dwarf temperature scale of \citet{sk82}.
For spectral types later than M0, \citet{luh99} developed a temperature 
scale for use with the evolutionary models of BCAH98 with $l_{mix}/H_p=1.0$ 
and 1.9 at $M/M_{\odot}\leq0.6$ and $M/M_{\odot}>0.6$, respectively.
This temperature scale was designed such that the members of the young 
quadruple system GG~Tau, which spanned types of K7 through M7.5, were coeval 
when placed on the BCAH98 model isochrones. 
On this scale, a spectral type of M7.5 corresponded to a temperature that
was roughly an average of the values for a dwarf and a giant.  
To extend the scale to M8 and M9, \citet{luh99} followed this trend and 
arbitrarily assigned temperatures that were intermediate between dwarf and 
giant scales. 
We now revise the temperature conversion from \citet{luh99} in the following 
manner. From the new surveys for low-mass members of IC~348 (this work) 
and Taurus \citep{bri02,luh03}, we have membership lists 
that are reasonably well-populated down to M9. As with the members of GG~Tau, 
the sequences for IC~348 and Taurus on the H-R diagram form empirical 
isochrones that can be used in defining a temperature scale that is compatible
with the models of BCAH98 and CBAH00.
When the populations of these two regions are placed on the H-R diagram with
the temperature scale from \citet{luh99}, they are parallel to the 
isochrones of BCAH98 and CBAH00 down to the latest spectral type present 
in GG~Tau (M7.5).
At M8 and M9, we now adjust the temperatures from \citet{luh99} so that 
the sequences for IC~348 and Taurus continue to fall parallel to the 
isochrones. The H-R diagrams of Taurus and IC~348 with this scale are 
plotted in the next section.
The revised temperature scale is tabulated in Table~\ref{tab:scale}
and is illustrated in Figure~\ref{fig:scale}. 
The temperature conversion is likely to be inaccurate at some level,
but because it falls between the scales for dwarfs and giants, the errors
are probably modest. However, regardless of the sizes of the errors and 
the true temperature scale for young objects, the scale we have described 
is an appropriate choice for use with the models of BCAH98 and CBAH00 
because this combination of scale and models is consistent with the available
observational constraints.

For reasons described in previous studies (e.g., \citet{luh99}), $I$ and $J$
are the preferred bands for measuring bolometric luminosities of young objects.
The adopted bolometric corrections for spectral types earlier than M9 are
described in \citet{luh99}. The components of four binaries with separations 
of 1-$3\arcsec$ (12AB, 42AB, 99AB, 259AB) are better resolved in the optical 
data. For the eight stars in these systems, the luminosities 
are computed by combining the bolometric corrections, the dereddened $I$-band 
measurements, and a distance modulus of 7.5. For all other objects, we use
$J$ instead of $I$. 
The extinctions, adopted spectral types, effective temperatures, and bolometric
luminosities are listed in Table~\ref{tab:mem}. 
Additional comments on the estimates of extinctions and luminosities for
individual sources are in \S~\ref{sec:append1b}.

\subsection{H-R Diagram}
\label{sec:hr}

\begin{figure}
\epsscale{0.7}
\plotone{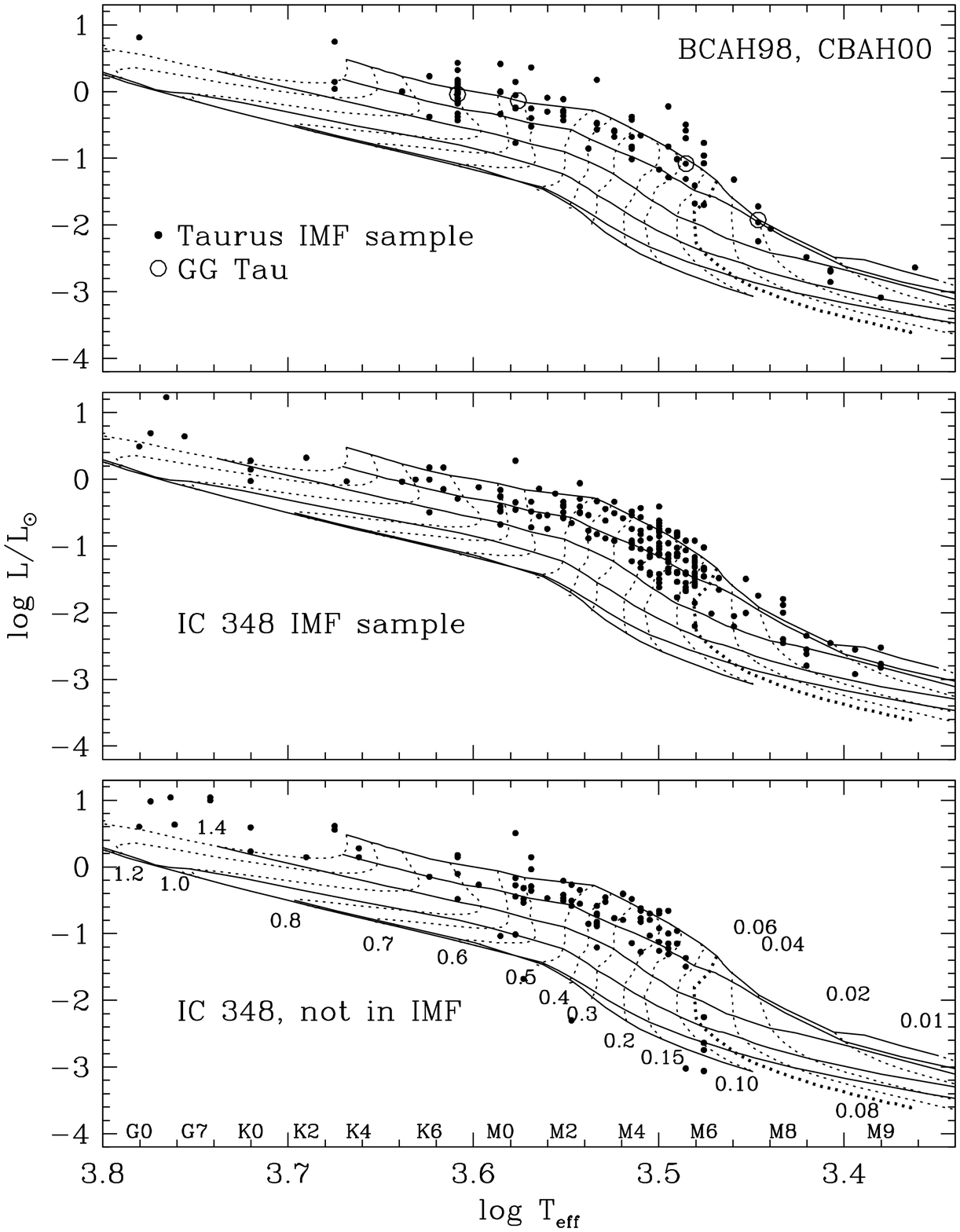}
\caption{H-R diagrams at low masses for objects in the IMFs for Taurus 
\citep{bri02,luh03} and for IC~348 (this work) ({\it top and middle}) that 
are shown in Figure~\ref{fig:imf}. These IMF samples are extinction-limited 
($A_V\leq4$) and apply to 8.4~deg$^2$ in Taurus and to the 
$16\arcmin\times14\arcmin$ field in IC~348 in Figure~\ref{fig:map}.
Members of IC~348 that are beyond the extinction threshold of $A_V=4$ 
or that have anomalously low luminosities for their spectral types are 
not included in the IMF ({\it bottom}). 
The latter sources are probably detected primarily in scattered light.
The theoretical evolutionary models of BCAH98 and CBAH00 are shown, where
the horizontal solid lines are isochrones representing ages of 1, 3, 10, 30,
and 100~Myr and the main sequence, from top to bottom.
The M spectral types have been converted to effective temperatures with a 
scale such that GG~Tau~Ba and Bb fall on the same model isochrone as
Aa and Ab and that the M8-M9 members of Taurus and IC~348 have 
model ages that are similar to those of the earlier members
(Table~\ref{tab:scale}, Figure~\ref{fig:scale}).
}
\label{fig:hr2}
\end{figure}
\begin{figure}
\epsscale{0.85}
\plotone{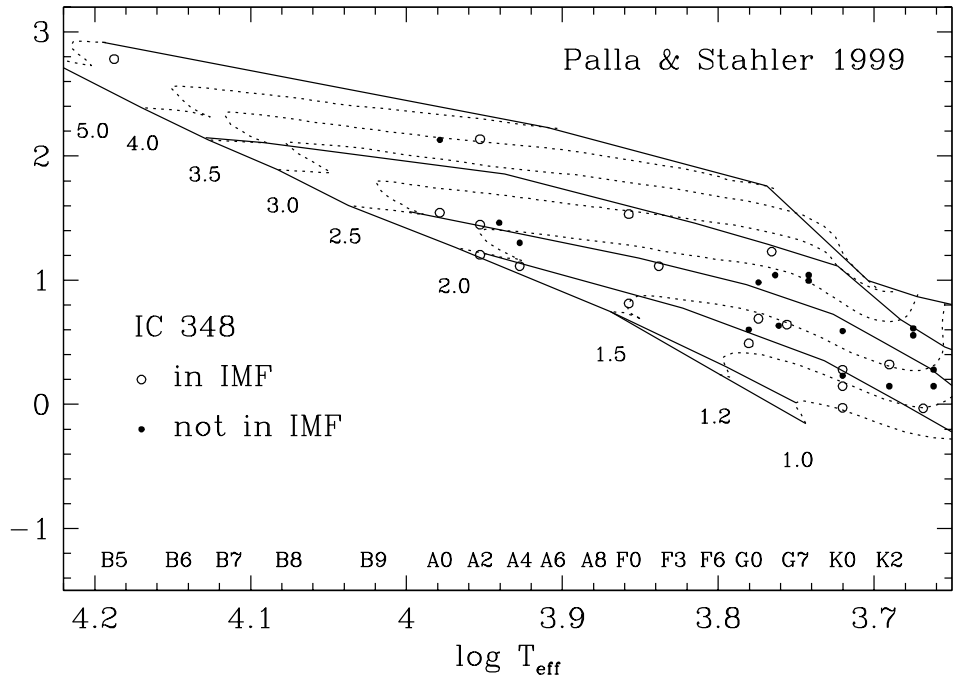}
\caption{
H-R diagram at high masses for members of IC~348. 
The theoretical evolutionary models of \citet{pal99} are shown, where 
the horizontal solid lines are isochrones representing ages of 0.1, 1, 3, 10, 
and 30~Myr and the zero-age main sequence, from top to bottom.
}
\label{fig:hr3}
\end{figure}

The effective temperatures and bolometric luminosities for the members of 
IC~348 can be converted to masses and ages with theoretical evolutionary 
models. Several sets of models for young, low-mass stars have been published in 
recent years \citep{bur97,dm97,bar98,pal99,cha00,sie00}, among which there are 
large differences in the predicted paths of young objects on the H-R diagram.
In \S~\ref{sec:append2}, we examine various observational tests of the 
available models at young ages to select the best calculations for 
interpreting the data in IC~348. In summary, 
the evolutionary models of \citet{pal99} are computed for masses of 
0.1-6~$M_\odot$ and agree well with dynamical mass estimates of young
stars at 1-6~$M_\odot$. Meanwhile, the calculations of BCAH98 and CBAH00 
consider masses of 0.001-1.4~$M_\odot$ and provide the best agreement among 
available models with observational constraints below a solar mass.
Therefore, for the subsequent analysis in this paper, we use the models
of \citet{pal99} for $M/M_\odot>1$ and the models of BCAH98 and
CBAH00 for $M/M_\odot\leq1$. Since the mass tracks of \citet{pal99}
and BCAH98 are similar near a solar mass, a continuous mass function can be 
derived for the full range of masses in IC~348. 
The low- and high-mass members of IC~348 are plotted with the evolutionary 
models on the H-R diagrams in Figs.~\ref{fig:hr2} and \ref{fig:hr3}. 
We differentiate between members that are included and excluded from the 
sample from which the IMF is generated (\S~\ref{sec:imf348}).
In Figure~\ref{fig:hr2}, we also show the sources in the IMF sample for Taurus
from \citet{luh03} and the components of the quadruple system GG~Tau.

As with previous H-R diagrams of populations in star-forming regions,
the sequences for Taurus and IC~348 exhibit finite widths that could 
correspond to distributions of ages within each region. 
For IC~348, the data would seem to imply ages ranging from 1 to 10~Myr.
However, there are several other potential sources of the observed thickness
in a cluster sequence, such as extinction uncertainties, unresolved binaries, 
variability from accretion and from rotation of spotted surfaces, and
differences in distances to individual members in extended regions like Taurus.
As a result, it is difficult to confidently measure star formation histories 
from H-R diagrams of star-forming regions \citep{kh90,har01} and it is unclear 
whether a spread of ages is actually reflected in the data for IC~348.

As long as unresolved binaries are not the dominant component of the
width of the cluster sequence, the median of the sequence should be a reflection
of the median age of the population, which is a useful characteristic age 
that can be compared among young populations. In Figure~\ref{fig:hr2}, the 
models indicate median ages of about 1 and 2~Myr for Taurus and IC~348, 
respectively. It is unlikely that errors in the distances to these regions 
are responsible for these apparent
age differences since isochrones of 1 and 2~Myr differ by one magnitude in 
luminosity, which is larger than the uncertainties in the distance moduli.
These relative ages for Taurus and IC~348 are consistent with the relative
evolutionary stages of disks in these regions \citep{kh95,hai00,hai01}.

We briefly describe nine objects in IC~348 that are anomalously faint for their
spectral types or colors, and therefore are possibly observed in scattered 
light. The first object, source 51, falls below the boundary separating 
candidate members from probable field stars in the diagram of $R-I$ versus $I$ 
from Luhman (1999). 
The presence of $K$-band excess emission and a red, featureless $K$-band 
spectrum are indicative of a star with a class~I spectral energy distribution
\citep{luh98b}.
This star also exhibits emission in X-rays and is surrounded by reflection 
nebulosity in our optical images. 
Eight sources fall near or below the main sequence on the H-R diagram in 
Figure~\ref{fig:hr2}.
Sources 203, 228, 276, 435, 621, 622, and 725 exhibit strong emission in 
H$\alpha$. Objects 203 and 435 show emission in forbidden lines. These two 
stars and object 276 also were detected in X-rays by \citet{pre01}.
Source 1434 appears to have strong H$\alpha$ emission and weak Na~I absorption,
each of which are indicative of youth, but given the low signal to noise 
of the spectrum, there is a small possibility that this star is a background
field dwarf. If this is the case, it must be close to the opposite side of the
cluster by its location on the H-R diagram.
With the possible exception of 1434, these sources exhibit photometry and 
signatures of youth that are consistent with young stars occulted by 
circumstellar structures, resulting in their detection primarily in scattered 
light.

Finally, the H-R diagram of the most massive members of IC~348
provides constraints on the distance of the cluster.
In Figure~\ref{fig:hr3}, the presence of cluster members 
on or near the zero-age main sequence indicates that IC~348
cannot be much closer than the adopted distance modulus of 7.5.
The recent analysis of $\delta$~Scuti-like pulsations in source 4 
by \citet{rip02} also supports this distance. 

\subsection{Initial Mass Function}
\label{sec:imf}

\subsubsection{The IC~348 Sample}
\label{sec:imf348}

We now measure the IMF for the $16\arcmin\times14\arcmin$ region in IC~348
shown in Figure~\ref{fig:map}. From the members in this field, we attempt to
construct a sample that is an accurate reflection of the cluster population.
In particular, the sample must be unbiased in mass so that the resulting
IMF is a meaningful representation of the cluster. For the IMF sample, we begin
by selecting all members that are in the $16\arcmin\times14\arcmin$ field and 
that have extinctions of $A_V\leq4$, which is high enough to include a large 
number of members while low enough that the completeness limit reaches low 
masses. As demonstrated in \S~\ref{sec:complete}, the current census of members 
in this field in IC~348 is unbiased in mass down to 0.03~$M_\odot$ for
$A_V\leq4$. The extinctions used in creating the extinction-limited 
sample are those listed in Table~\ref{tab:mem}. 
The anomalously faint stars listed in \S~\ref{sec:hr} that fall near 
or below the main sequence in Figure~\ref{fig:hr2} are rejected from the 
IMF sample because the membership census is not complete for objects of this
type (\S~\ref{sec:complete}).

There are five sources in Table~\ref{tab:mem} that lack measurements of
extinctions and luminosities. Should any of these objects be included in the
IMF sample? 
The spectra and IR colors of sources 46 and 51 imply extinctions that are
higher than the limit defining the IMF sample. The other three stars, 60B, 78B, 
and 187B, are secondaries that lack measured spectral types. Because 60A and 
187A have extinctions that place them in the IMF sample, we include their 
companions as well. To estimate the masses of 60B and 187B, we assume that the 
components of each binary have the same reddenings and ages and that the 
ratios of their luminosities are equal to the ratios of the fluxes at
$I$ and $Z$ for 60B (this work) and at $H$ for 187B \citep{duc99}.
By combining these assumptions with the evolutionary models,
we arrive at masses of 0.34 and 0.1~$M_\odot$ for 60B and 187B.

After applying the above criteria, the IMF sample contains 194 objects. 
With the exception of 60B and 187B, these sources are
indicated in the H-R diagrams in Figs.~\ref{fig:hr2} and \ref{fig:hr3}.
The masses for these objects are inferred from the choice of theoretical 
evolutionary models described in the previous section.

\subsubsection{Previous Studies of IC~348}

Our measurement of the IMF in IC~348 should be placed in the context of 
previous work on this cluster. We have derived an IMF for IC~348 
by combining the positions of cluster members on 
the H-R diagram with evolutionary models. The sample includes all known 
members with $A_V\leq4$ in a $16\arcmin\times14\arcmin$ region of the cluster
and is unbiased in mass for $M/M_\odot\geq0.03$. 
\citet{her98}, \citet{luh98b}, and NTC00 also used H-R diagrams and 
evolutionary models to infer mass functions. \citet{her98} presented an IMF 
that contained all probable members in a $14\arcmin\times8\arcmin$ field and
that reached down to masses of $\sim0.2$~$M_\odot$. The IMF from \citet{luh98b}
for the $5\arcmin\times5\arcmin$ center of IC~348 was complete 
for $M/M_\odot\gtrsim0.1$ and extended to $0.02$~$M_\odot$. Because
all of the membership information and other relevant data from \citet{her98}
and \citet{luh98b} are included in our analysis and because of
our improved methods of interpreting the data (e.g., choice of evolutionary 
models), we do not compare our IMF to the results of those studies.

Studies by \citet{mue03} and \citet{tej02} have arrived
at IMFs for IC~348 through analysis of broad-band photometry. 
\citet{mue03} obtained deep IR photometry for the entire 
cluster and estimated an IMF from the resulting luminosity functions.
The IMFs from \citet{mue03} and this work are generally consistent 
with each other. For instance, both IMF measurements exhibit peaks near
0.1-0.2~$M_\odot$. The uncertainties in the IMF of \citet{mue03} are
largest below the hydrogen burning limit because of the rapidly increasing
contamination by background stars at the fainter levels of the IR luminosity
functions. Meanwhile, \citet{tej02} used optical and IR photometry from the 
Guide Star Catalog and 2MASS to identify candidate members within a radius of
$20\arcmin$ from the cluster center. They estimated masses for these candidates
by combining $K$ photometry, evolutionary models, and the mean age and 
extinction for cluster members. We find several problems in the analysis
of \citet{tej02}. First, the assumption of fixed values for extinction 
and age for the cluster members is overly simplified, as demonstrated by the
scatter in the sequence for IC~348 on the H-R diagram and the range of 
extinctions that we derive. Second, to correct for
field star contamination in their sample of candidate cluster members, they 
subtracted data from off-cluster control fields. However, to properly 
correct for field stars in this way, one must estimate the
distribution of extinctions toward background stars in the cluster field,
apply extinction to the data from the control field to duplicate that
reddening distribution, and then perform the subtraction.
\citet{tej02} did not apply an extinction model to their control field,
which is especially necessary when optical data are involved. Finally,
their IMF sample, like that of NTC00 later in this section,
is susceptible to a bias against low-mass sources because it includes all 
candidate members rather than only those within an extinction threshold.

In the empirical H-R diagram constructed from the narrowband steam measurements,
NTC00 assumed that objects appearing near the cluster sequence were members.
From these sources, they derived an IMF that was reported to be
complete to the deuterium burning limit ($K\sim16.5$ for $A_V=3$,
0.013-0.015~$M_{\odot}$; \citet{bur97}).
How do our findings in \S~\ref{sec:najita} regarding the membership and 
spectral types of the NICMOS sources affect the IMF reported by NTC00?
At $K>14$, photometric uncertainties in the NICMOS narrow-band photometry 
translated into very large errors in the estimated spectral types. Consequently,
several field stars were mistaken for objects with spectral types of late M. 
In addition, because the spectral types from the steam data were
systematically too late by 1-2 subclasses, the masses were underestimated. 
An offset of one subclass in the substellar
regime corresponds to a change of a factor of $\sim2$ in the inferred mass.
These two effects both resulted in an overestimate of the number of 
brown dwarfs by NTC00. In Figure~\ref{fig:comp},
28 sources have NICMOS classifications later than M6, when only 9 of 
these sources actually have optical spectral types in that range. 
NTC00 concluded that their spectral types were accurate down to
$K=16.5$, or the brightness of objects at the deuterium burning mass limit 
for the average age and extinction of cluster members (3~Myr, $A_V=3$).
This mass was thus quoted as the completeness limit of their IMF for
the NICMOS fields in IC~348. 
However, if an IMF sample includes all probable members, as in NTC00,
then the mass completeness limit is the lowest mass at which 
the sample is complete to the {\it maximum} age and extinction. 
To be representative of the population down to the deuterium burning limit,
an IMF from the data of NTC00 would need to contain only sources
with ages and reddenings below the mean values. In addition, it is not clear 
that their spectral classifications are sufficiently accurate down to $K=16.5$ 
for mass estimates and for the measurement of an IMF;
the average uncertainties for the spectral types reported by NTC00
are $\pm3.5$ subclasses for types later than M2 with $K=14.5$ to 
16.5. Indeed, in \S~\ref{sec:najita} we showed that at least two objects at 
$K<16.5$ with NICMOS classifications later than M6 are field stars.
Also, because of the large uncertainties in both their spectral types and 
dereddened $K$ magnitudes, the cluster sequence overlaps with the field
star population (Figure~\ref{fig:nic}). Without a clear separation of the two 
populations or other membership information, the low-mass members cannot 
be reliably identified and added into the IMF. Given these various issues,
the IMF from NTC00 is probably not accurate in the substellar regime or 
complete to the deuterium burning mass limit. 

\subsubsection{Comparison of IMFs in Different Environments}

\begin{figure}
\epsscale{0.75}
\plotone{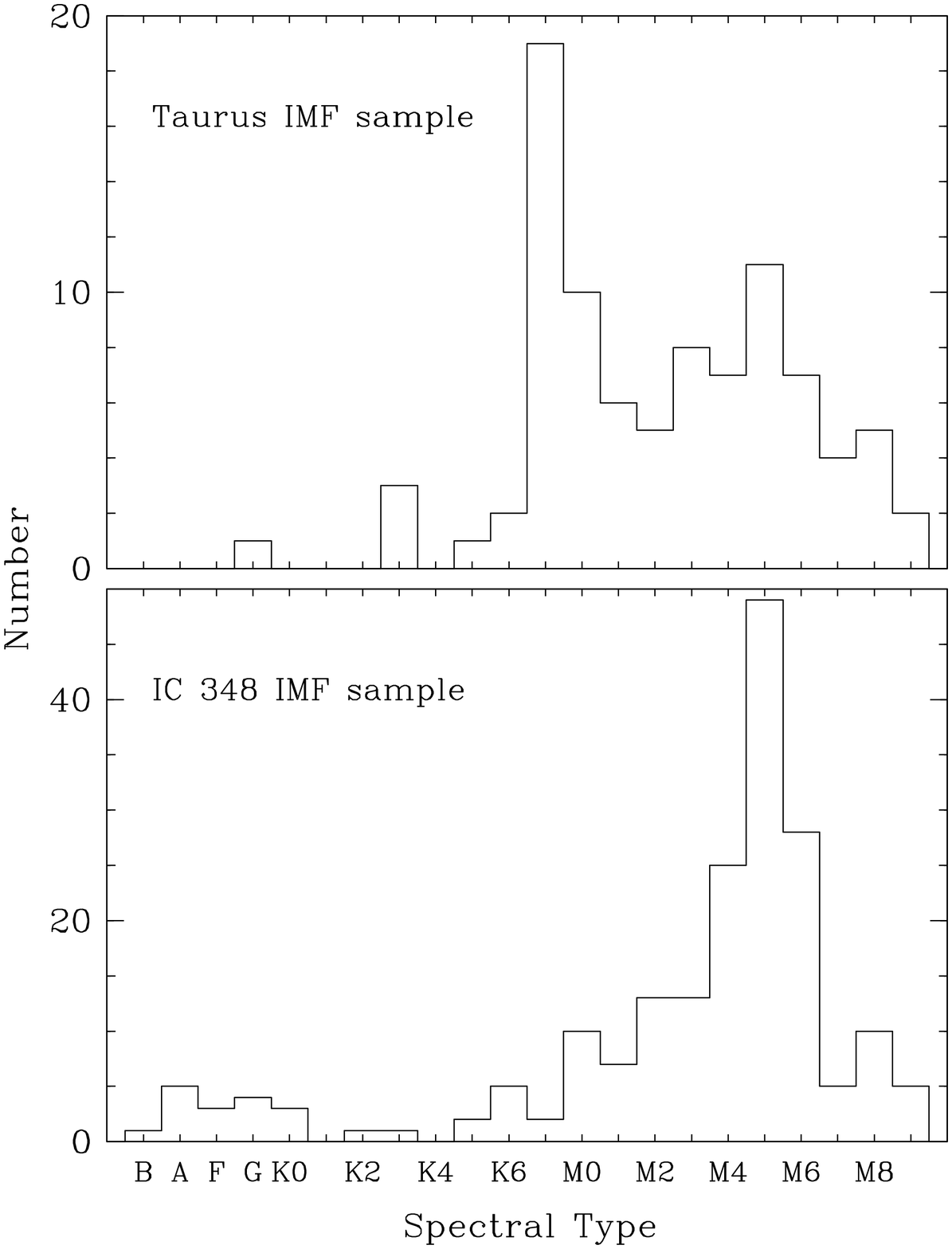}
\caption{Distributions of spectral types for objects in the IMFs 
for Taurus \citep{bri02,luh03} and for IC~348 (this work) that are shown 
in Figure~\ref{fig:imf}. 
These samples are extinction-limited ($A_V\leq4$) and apply to 8.4~deg$^2$ in 
Taurus and to the $16\arcmin\times14\arcmin$ field in IC~348 in 
Figure~\ref{fig:map} and are nearly 100\% complete for spectral types of 
$\leq$M9 and $\leq$M8, respectively.
Because the evolutionary tracks for young low-mass stars are mostly
vertical, spectral types should be closely correlated with stellar masses.
As a result, these distributions of spectral types should directly reflect the 
IMFs in IC~348 and Taurus. This comparison provides clear, model-independent 
evidence for significant differences in the IMFs of these two regions.
}
\label{fig:histo1}
\end{figure}

By using methods similar to those in this study, \citet{luh00a}, \citet{bri02},
and \citet{luh03} surveyed 8.4~deg$^2$ in the Taurus star-forming region and 
arrived at a census of members that is complete for $M/M_\odot\geq0.02$ and 
$A_V\leq4$.  \citet{luh03} presented an IMF from the 92 Taurus members within 
this extinction threshold and in the survey fields. We now compare that IMF 
sample for Taurus to the one that we have defined for IC~348.
Because of the vertical nature of the mass tracks for low-mass stars on the H-R 
diagram, the spectral types of young objects should be well-correlated with 
their masses. Very little evolution in temperature is expected between the 
ages of Taurus and IC~348 (1 and 2~Myr), implying virtually identical
relations between spectral types and masses for these two populations.
In addition, a spectral type is a simple, observable quantity that can be 
measured to good accuracy with relative ease, particularly at M types.
Therefore, we use the distributions of spectral types for IC~348 and 
Taurus as IMF proxies that can be compared in a straightforward, reliable
fashion without the involvement of evolutionary models. 
The IMF samples for IC~348 and Taurus are unbiased in mass down to 0.03 and 
0.02~$M_\odot$, respectively, which correspond to types of $\sim$M8 and M9 
for young ages. 
One and two sources in the IMFs for Taurus and IC~348, respectively, lack
measured spectral types. After omitting these three stars, the numbers 
of objects as a function of spectral type in the IMF samples for IC~348 
and Taurus are plotted in Figure~\ref{fig:histo1}. 
The distribution for IC~348 reaches a maximum at M5, while primary and 
secondary peaks appear at K7 and M5 in Taurus. Spectral types of M5 and K7
correspond to masses near 0.15 and 0.8~$M_\odot$ for ages of a few million
years by the models of BCAH98.
The spectral type distributions for IC~348 and Taurus provide clear, unambiguous
evidence for a significant variation of the IMF with star-forming conditions.
Previous studies have noted that a difference of this kind might be present
between Taurus and clusters like Orion (e.g., \citet{hil97}).
However, most previous samples of 
members of star-forming regions have been derived by combining disparate 
surveys that utilized biased selection techniques (e.g., H$\alpha$, IR excess). 
It was unclear whether the predominance of K7 and M0 stars in Taurus was 
an accurate reflection of the region or simply a result of incompleteness at
later types. But because of the new magnitude-limited membership surveys in 
IC~348 and Taurus, we have been able to make the first comparison of spectral
type distributions in which the samples are complete down to late
spectral types and contain relatively large numbers of members.

\begin{figure}
\plotone{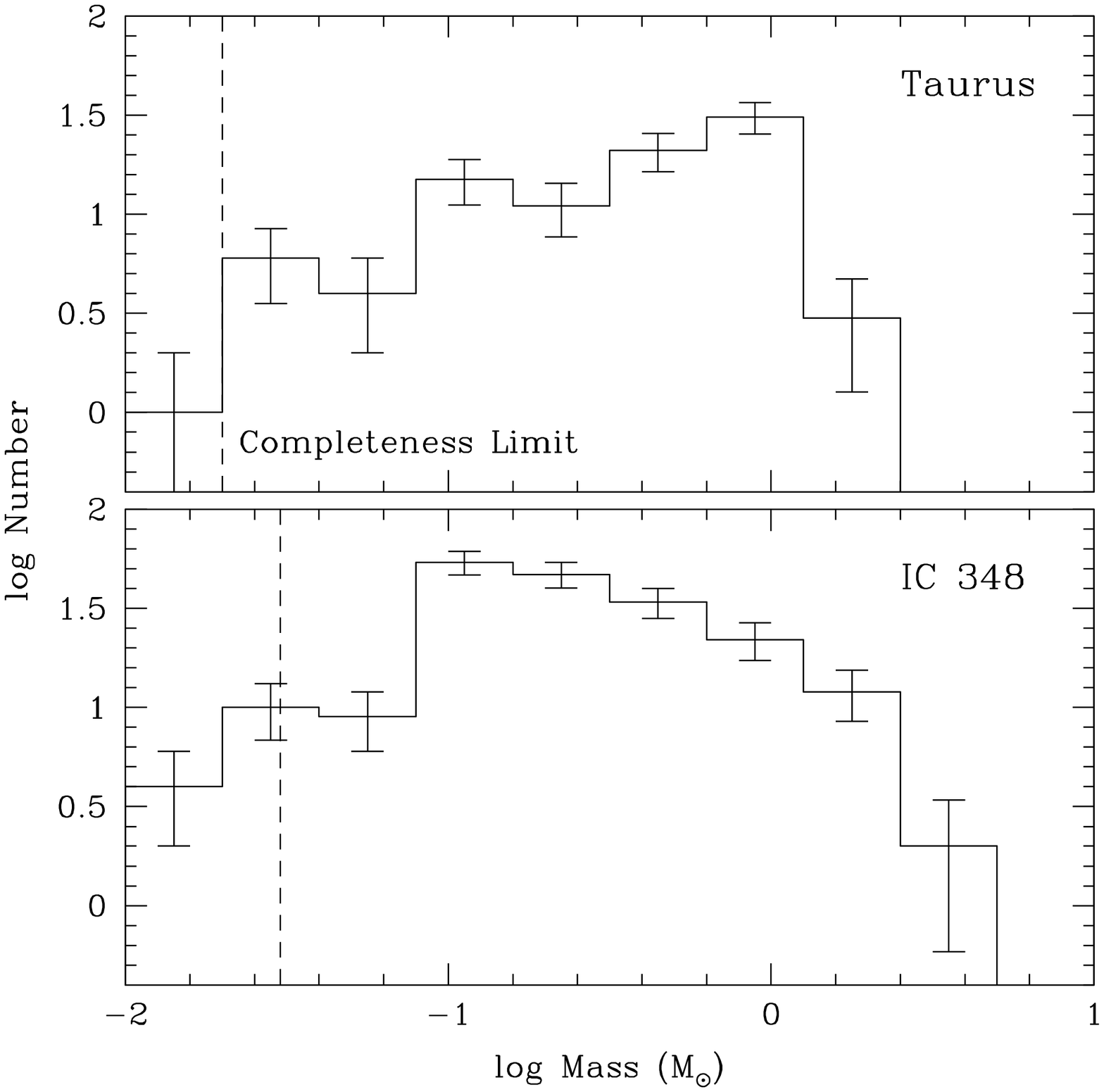}
\caption{
IMFs for extinction-limited samples ($A_V\leq4$) in 8.4~deg$^2$ in Taurus 
\citep{bri02,luh03} and in the $16\arcmin\times14\arcmin$ field in IC~348 in 
Figure~\ref{fig:map} (this work). These samples are unbiased in mass 
for $M/M_\odot\geq0.02$ and 0.03, respectively.
In the units of this diagram, the Salpeter slope is 1.35.
}
\label{fig:imf}
\end{figure}

When the data for the samples in IC~348 and Taurus are transformed
to individual masses with evolutionary models (\S~\ref{sec:imf348};
\citet{bri02,luh03}), the IMFs in Figure~\ref{fig:imf} are produced. 
Because the same techniques and evolutionary models were employed in converting
from data to masses for each population, we can be confident in the
validity of any differences we find in the IMFs for IC~348 and Taurus.
The IMF for Taurus peaks near 0.8~$M_\odot$ and steadily declines to lower
masses. While two peaks are present in the distribution of spectral types for 
Taurus, only one is found at a significant level in the IMF. Apparently,
the bimodal appearance of the spectral type distribution is simply the product
of stellar evolution convolved with the particular form of single-peaked IMF 
found in Taurus. Meanwhile, the mass function for IC~348 rises from high 
masses down to a solar mass in a roughly Salpeter fashion, rises more slowly 
down to a maximum at 0.1-0.2~$M_\odot$, and then declines into the substellar 
regime. We have quantified the significance of the differences in the 
distributions of spectral types and masses for IC~348 and Taurus by performing 
a two-sided Kolmogorov-Smirnov test between the distributions for spectral types
of $\leq$M8 and for masses of $M/M_{\odot}\geq0.03$. In terms of both spectral 
types and masses, the probability that the samples for IC~348 and Taurus are 
drawn from the same distribution is $\sim0.01$\%.

It is unlikely that the IMFs of single objects  -- both isolated sources and 
individual components of multiple systems -- in IC~348 and Taurus are the 
same and these observed differences are simply the result of dynamical effects, 
such as more frequent stripping of low-mass companions in the denser 
environment of IC~348. 
Companions at separations greater than $1\arcsec$ and $2\arcsec$ in IC~348 
and Taurus, respectively, were included in both IMFs. 
As a result, the disruption of wide binaries should have no effect on our
results as long as the components remain within the cluster, which is likely
for these loosely bound systems. 
In order for the spectral type distributions for all objects earlier than
M9 to be the same in Taurus and IC~348, enormous numbers of M stars would 
need to be hidden as close companions in Taurus ($\sim10$ per primary).
In addition, if the Taurus IMF is computed
for only the most compact aggregate, L1495E, which has a stellar density only
five times lower than that of IC~348, than the IMF variations between these
two regions persist. Finally, out of the various types of stellar populations 
in which mass functions are measured, star-forming clusters with low to 
moderate stellar densities like Taurus and IC~348 are the least likely 
sites to be affected significantly by dynamical evolution.
For instance, in IC~348, the crossing time of
a star moving at the escape velocity ($\sim0.8$~km~s$^{-1}$, \citet{her98})
is comparable to the age of the cluster (2~Myr). When the members of IC~348
were younger and less evolved, they likely had even lower velocities than they 
do now, as indicated by the protostellar clumps in this cluster \citep{bac87}. 

The shapes of the IMFs derived for Taurus and IC~348 are sensitive to the 
adopted temperature scale and evolutionary models. The combination used here 
was designed to produce the best agreement with the various 
observational constraints (\S~\ref{sec:append2}). In particular, the IMFs above 
$\sim0.5$~$M_\odot$ should be fairly accurate since our choices of temperature 
scale and models implies masses from the H-R diagram that are consistent with
the dynamical mass estimates of young stars. However, because fewer 
constraints are available at low masses, the IMFs below
$\sim0.5$~$M_\odot$ could be subject to systematic errors. 

Even with the extensive spectroscopic work that has been done in Orion, one
can define a well-populated sample of members that is complete down to only
mid-M types in that region. To extend the IMF in Orion to lower masses,
studies have relied on luminosity function modeling \citep{hc00,luh00b,mue02}.
As a result, a comparison of the IMFs in Taurus and IC~348 to that in Orion 
is less definitive than the comparison between the first two regions.
We consider the IMF reported by \citet{luh00b} for the Trapezium Cluster,
which was derived with similar methods to the ones used for Taurus and IC~348.
The IMF from \citet{luh00b} peaks near 0.6~$M_\odot$ and is roughly 
flat to lower masses, and thus differs somewhat from
the IMF for IC~348 in Figure~\ref{fig:imf}. It is unclear whether this 
difference is real or is the result of a shortcoming in the merging of the
spectroscopic data and the luminosity function modeling for the Trapezium.
Meanwhile, the IMFs for IC~348 and the Trapezium derived from luminosity 
function modeling are similar \citep{mue03}, but do allow for subtle 
variations of this type.
To reliably compare these IMFs at this level of detail, the completeness 
limit of the spectroscopic work in the Trapezium must be extended to later 
spectral types.

The data for the Trapezium and other young clusters are more amenable to 
a comparison of the global properties of the IMFs.
We define ratios to quantify the relative numbers of brown dwarfs and stars 
and the relative numbers of low-mass and high-mass stars:

$$ {\mathcal R}_{1} = N(0.02\leq M/M_\odot\leq0.08)/N(0.08<M/M_\odot\leq10)$$

$$ {\mathcal R}_{2} = N(1<M/M_\odot\leq10)/N(0.15<M/M_\odot\leq1)$$

In Table~\ref{tab:ratios}, we compare these ratios for IMFs in 
IC~348 (this work), Taurus \citep{bri02,luh03}, and the 
Trapezium \citep{luh00b} and in the open clusters of the Pleiades \citep{bou98} 
and M35 \citep{bar01}. The values for the latter three clusters were compiled
by \citet{bri02}. For Taurus, we update the ratios in \citet{bri02} by 
adding the new members from \citet{luh03}, producing 
${\mathcal R}_{1} = 11/80 = 0.14\pm0.04$ and 
${\mathcal R}_{2} = 5/61 = 0.08\pm0.04$. 
From the IMF for IC~348, we have 
${\mathcal R}_{1} = 21/169 = 0.12\pm0.03$ and 
${\mathcal R}_{2} = 18/99 = 0.18\pm0.04$. 
The former quantity could be slightly underestimated since the IC~348 sample
may be incomplete at 0.02-0.03~$M_{\odot}$ (\S~\ref{sec:complete}). 
Note that this measurement of ${\mathcal R}_{1}$ is consistent with the value
implied by the luminosity function modeling of \citet{mue03}.
As shown in Table~\ref{tab:ratios}, the relative numbers of high- and 
low-mass stars in IC~348 are similar to those found in the Trapezium
and in the open clusters, while the frequency of brown dwarfs appears to be
lower in IC~348 than in the Trapezium and near the value for Taurus. 
The IMF measured for the Ophiuchus star-forming cluster, which is comparable
to IC~348 in stellar density ($n=100$-1000~pc$^{-3}$), is consistent 
with that of IC~348 
\citep{lr99,luh00b}. However, because the number statistics are relatively
poor in the data for Ophiuchus, we do not include it in Table~\ref{tab:ratios}.

The implications of the variations in the IMF among Taurus, the Trapezium,
and the open clusters for models of the IMF were discussed by \citet{bri02}.
They suggested that the lower frequency of brown dwarfs in Taurus relative to
the Trapezium could reflect differences in the typical Jeans masses of
the two regions. This scenario continues to be plausible when the data for
IC~348 are included. IC~348 contains a much higher frequency of low-mass stars
(0.1-0.2~$M_{\odot}$) than Taurus, but a comparable fraction of brown dwarfs, 
which would imply an average Jeans mass for IC~348 that is intermediate between 
those of Taurus and the Trapezium. 
The observed IMF variations between Taurus and IC348 might also be explained
through the recent model of turbulent fragmentation by \citet{pn02}. In those
calculations, the IMF peaks at a higher mass for lower values of the gas
density and the Mach number. Indeed, Taurus exhibits a higher peak mass
than IC~348 and is less dense and more quiescent than most star-forming regions.
Meanwhile, these variations would seem difficult to 
explain with models in which the shape of the IMF is determined by the 
competition between accretion and outflows \citep{af96}.
The significant variation that we find in the IMF, which is particularly 
well-established between Taurus and IC~348, comprises a new and important test 
for any model of the IMF.

\subsection{X-ray Properties}
\label{sec:xray}

The X-ray properties of young stars in IC~348 have been described 
by \citet{pre96}, \citet{pre01}, and \citet{pre02}.
We update the results of those studies with the data from our survey for new
members.

\begin{figure}
\plotone{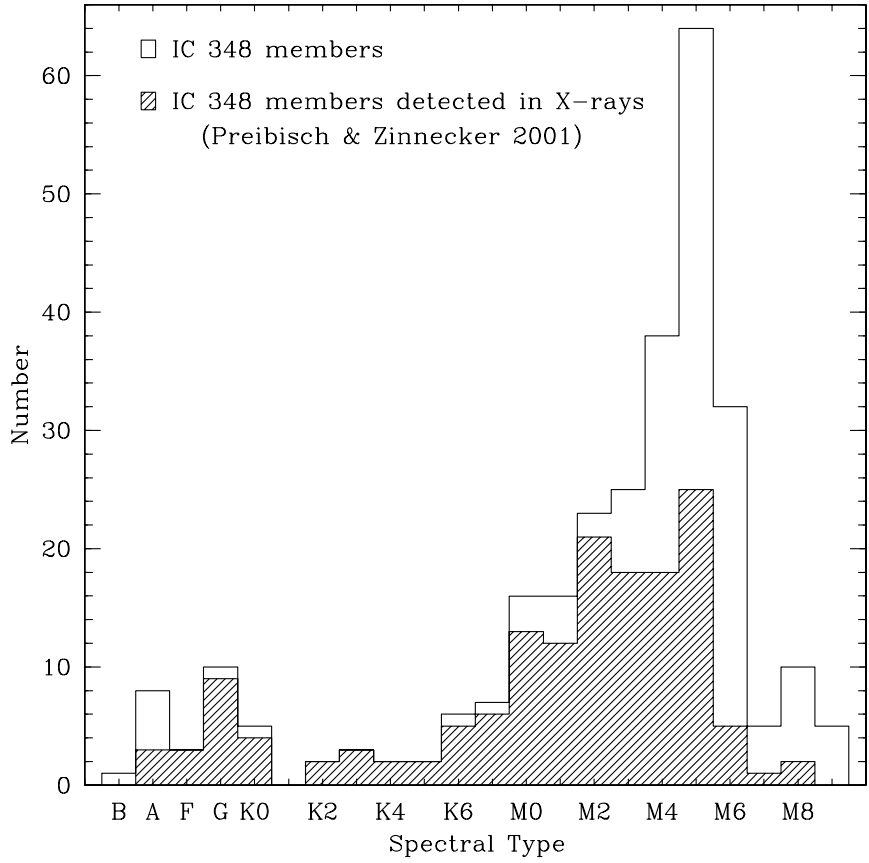}
\caption{
Distribution of spectral types for the 283 members of IC~348 that have 
accurate classifications. \citet{pre01} have detected X-rays from 154 of these 
sources, as indicated by the shaded histogram.
}
\label{fig:histo2}
\end{figure}

The deepest X-ray observations of IC~348 to date were conducted by 
\citet{pre01} with the {\it Chandra X-ray Observatory}.
In a $17\arcmin\times17\arcmin$ field toward the cluster, 
they detected 219 sources, which included 119 previously known members
and 58 potential new members.  
Among the 283 known members of IC~348 that have measured spectral types,
154 members are detected in X-rays. The fraction of members with observed X-ray 
emission as a function of spectral type is illustrated by the distributions
of spectral types in Figure~\ref{fig:histo2}.
Source 51, which lacks a spectral type measurement and is a probable 
class~I object, is an X-ray source as well.
We now examine the nature of the remaining 64 X-ray detections.
Most of the 39 X-ray sources that lack counterparts in available optical 
and IR data are probably in the background of IC~348 and unrelated to the 
cluster \citep{pre01}. One X-ray source, object 77, is identified as 
a field star through proper motions measurements (\S~\ref{sec:mem}).
Ten objects are candidate members, one of which is within the 
$16\arcmin\times14\arcmin$ field in Figure~\ref{fig:map} and thus appears in 
Figure~\ref{fig:col}. The 12 X-ray sources that are below the boundary are 
probably in the foreground or background of the cluster. It is possible that
a few of these sources could be cluster members observed in scattered light,
which would make them appear subluminous on a color-magnitude diagram, but
none of these objects show any other evidence of youth or membership.
Finally, two X-ray sources are detected only in IR data.
\citet{pre01} identified one of these sources, 
2MASS~03444330+3201315, as a possible class~I object by its strong $K$-band
excess emission and high extinction. At the position of this object 
in our $I$ and $Z$ images, we detect reflection nebulosity extended across 
$15\arcsec$ and no point source. The IR counterpart to the second X-ray source
is 2MASS~03441977+3159190, which has been detected only at $K_s$.
Our optical images show faint nebulosity in the vicinity of this object's 
position. If 2MASS~03441977+3159190 is a member of IC~348, it must be highly 
embedded, probably at the class~I stage. Both 2MASS~03444330+3201315 and
2MASS~03441977+3159190 are in the southern part of IC~348 where the cluster
merges into the Perseus molecular cloud and where the most embedded and least 
evolved known members of the cluster reside, such as sources 51 and 13 
(IC~348-IR) and a likely class~0 object \citep{mcc94}. 

\citet{pre01} referred to several objects in IC~348 
as brown dwarfs or brown dwarf candidates because of the late spectral types 
implied by the steam measurements of NTC00.
However, most of these sources have optical spectral types that are earlier 
that M6 and therefore are probably low-mass stars rather than brown dwarfs 
(\S~\ref{sec:najita}). There are three known members of IC~348 (329, 355, 613) 
that are likely brown dwarfs ($>$M6) and that have been detected in X-rays.

\section{Conclusions}

The results of our new census of young stars and brown dwarfs in the IC~348 
star-forming cluster are summarized as follows:

We have obtained deep images at $I$ and $Z$ for a $42\arcmin\times28\arcmin$
field encompassing the IC~348 cluster.
By combining these data with optical and IR photometry from previous
surveys, we have constructed extinction-corrected color-magnitude diagrams and
used them to select candidate members.
Through spectroscopy of these candidates, we have identified 122 new 
members, 15 of which have spectral types of M6.5-M9, corresponding to masses
of $\sim0.08$-0.015~$M_\odot$ by the evolutionary models of BCAH98 and CBAH00.
After examining the membership status of all objects toward IC~348 that
have been observed spectroscopically in this work and in previous studies, we 
have arrived at a list of 288 known members of IC~348, 23 of which are later
than M6 and thus are likely to be brown dwarfs.
We find a lower proportion of such late-type members than in the study of 
NTC00 because of a systematic calibration error in the translation 
of their photometric data to spectral types, and their identification of a
number of faint background stars as low-mass cluster members.

We have estimated extinctions, luminosities, and effective temperatures for
the known members of IC~348 and have placed these sources on the H-R diagram.
To select the best calculations with which to interpret these data, we have 
compiled observational constraints from previous studies \citep{whi99,sim00} 
and from this work on IC~348 and have applied them to the available 
evolutionary models.
These tests tend to support the validity of the models of \citet{pal99}
at $M/M_\odot\gtrsim1$ and the models of BCAH98 and CBAH00 at 
$M/M_\odot\lesssim1$. We have combined these models with the
H-R diagram for IC~348 to infer masses for individual members of the cluster. 
For a $16\arcmin\times14\arcmin$ field centered on IC~348, we have defined 
an extinction-limited sample of known members ($A_V\leq4$) that is unbiased in 
mass and nearly 100\% complete for $M/M_{\odot}\geq0.03$ ($\lesssim$M8).
In logarithmic units where the Salpeter slope is 1.35, the IMF for
this sample in IC~348 rises from high masses down to a solar 
mass, rises more slowly down to a maximum at 0.1-0.2~$M_\odot$, and then 
declines into the substellar regime. In comparison, the 
similarly-derived IMF for Taurus from \citet{luh03} rises quickly to 
a peak near 0.8~$M_\odot$ and steadily declines to lower masses.
The distinctive shapes of the IMFs in IC~348 and Taurus are reflected in the 
distributions of spectral types, which peak at M5 and K7, respectively.
This is the first comparison of spectral type distributions between two 
star-forming populations that is based on samples that are complete to late
spectral types and that include relatively large numbers of members, and it
represents clear, model-independent evidence for a significant variation 
of the IMF with star-forming conditions.

\acknowledgements

We thank Perry Berlind and Mike Calkins for performing the FAST observations
and Jean-Charles Cuillandre for reducing the CFH12K images.
We are grateful to Isabelle Baraffe and Francesco Palla for access to their
most recent calculations and to Lee Hartmann for comments on the manuscript.
K. L. was supported by grant NAG5-11627 from the NASA Longterm Space
Astrophysics program.
The $JHK$ data from \citet{mue03} were obtained with FLAMINGOS under the NOAO 
Survey Program ``Towards a Complete Near-Infrared Imaging and Spectroscopic 
Survey of Giant Molecular Clouds" (PI: E. Lada) and supported by NSF grants
AST97-3367 and AST02-02976 to the University of Florida. FLAMINGOS was
designed and constructed by the IR instrumentation group
(PI: R. Elston) at the Department of Astronomy at the University of Florida
with support from NSF grant AST97-31180 and Kitt Peak National Observatory.
This research has made use of the NASA/IPAC Infrared Science Archive, which is
operated by the Jet Propulsion Laboratory, California Institute of Technology, 
under contract with the National Aeronautics and Space Administration. 
Some of the data presented herein were obtained at the MMT Observatory,
a joint facility of the Smithsonian Institution and the University of Arizona.
Data were also obtained at the W. M. Keck Observatory, 
which is operated as a scientific partnership among the California Institute 
of Technology, the University of California, and the National Aeronautics and 
Space Administration. The Observatory was made possible by the generous 
financial support of the W. M. Keck Foundation.
We wish to extend special thanks to those of Hawaiian ancestry on whose sacred
mountain we are privileged to be guests. Without their generous hospitality,
some of the observations presented herein would not have been possible.

\appendix

\section{Comments on Individual Sources}
\label{sec:append1}

\subsection{Spectral Types and Membership}
\label{sec:append1a}

As explained in \S~\ref{sec:sptype}, our optical spectra do not differentiate
well between giants, dwarfs, and pre-main-sequence objects at G and early K 
types. We have used IR spectra to break this degeneracy in the classifications
of several objects in our sample. For another of these stars, source 44, we
measured an equivalent width of $0.35\pm0.05$~\AA\ for Li at 6707~\AA. Because 
$\sim99$\% of field giant exhibit Li strengths below 0.1~\AA\ \citep{bro89}, 
this star is unlikely to be a background giant. Meanwhile, it cannot be a 
foreground dwarf given the significant reddening in its colors and spectrum. 
Therefore, we take object 44 to be a cluster member. 
Five of the remaining sources with ambiguous optical classifications 
(20, 22, 47, 53, 79) were detected in the X-ray observations of \citet{pre01}.
By their reddened colors, these stars must be either 
members of IC~348 or background giants rather than foreground stars.
All of these stars have $L_{\rm X}/L_{\rm bol}>10^{-4}$, whereas 
most G and K giants have $L_{\rm X}/L_{\rm bol}<10^{-7}$ and only a few
have been observed with ratios as high as $10^{-4}$ \citep{hue96}.
Therefore, these five stars are probably members of the cluster.

One of the stars with ambiguous optical types, source 81, lacks Li and IR 
spectroscopic measurements and is not detected in X-rays. 
The membership status is 
uncertain for this star and for two other stars in Table~\ref{tab:unc}.
Source 197 is highly reddened and falls between the 30~Myr isochrone and
the main sequence when placed on the H-R diagram for the distance of IC~348. 
This star could be a member of IC~348, possibly observed in scattered light,
or a field dwarf that is near the opposite side of the cluster.
Because source 1927 exhibits little or no extinction in its spectrum and colors 
and shows no other evidence of youth, it could be either a cluster member or 
a foreground M dwarf.

In a spectrum with low signal-to-noise, source 1476 appears to have strong 
emission in H$\alpha$ and strong absorption in Na~I, which are suggestive of
a young star and a field dwarf, respectively. Given its position below the
main sequence on the H-R diagram for the distance of IC~348, this star is
probably a background dwarf. 
For star 404, the presence of dwarf-like Na~I absorption and the lack of 
reddening or evidence of youth are indicative of a foreground field dwarf.
Because the spectrum of object 906 is matched better with an average of a 
dwarf and a giant than with a dwarf, we take it to be young and thus a
cluster member. However, the low signal-to-noise of the data for this
source precludes a reliable measurement of the gravity-sensitive Na~I and
K~I features. In addition, this source lacks significant extinction or
emission lines. As a result, there is a small possibility that this source 
is a foreground dwarf.

\subsection{Luminosities}
\label{sec:append1b}

The most massive members of IC~348 are the components of the B5 binary system
BD+$31\arcdeg$643 (source 1 in Table~\ref{tab:mem}), which have 
$\Delta K=0.16$ and a separation of $0\farcs6$ \citep{kal97}.
We computed a luminosity from the unresolved photometry of the 
system and divided the result by two for Table~\ref{tab:mem} and 
Figure~\ref{fig:hr3}.
When estimating the luminosities for 60A, 78A, and 187A from the $J$-band 
photometry (\S~\ref{sec:teff}), we corrected for the contribution of the 
secondaries by assuming that $\Delta J$ was equal to
$\Delta I, Z=1.1$ (this work), $\Delta K=2.65$ (NTC00),
and $\Delta H=0.99$ \citep{duc99}, respectively.
For sources 12A, 12B, 42A, 42B, 99A, and 99B, we measured the luminosities from 
our resolved $I$-band photometry.

\section{Evolutionary Models}
\label{sec:append2}

\subsection{Previous Tests}

To evaluate the validity of the available evolutionary models at young ages, we 
first review the previous observational tests that have been applied to them.

The potential coevality of components of young multiple systems can be used as
a test of evolutionary models \citep{har94,pra98}.
\citet{whi99} obtained accurate, resolved photometry and spectroscopy
for the members of the young quadruple system GG~Tau, which span spectral
types from K7 to M7.5.
They placed these sources on the H-R diagram for dwarf and giant temperature
scales and compared the resulting empirical isochrone to isochrones from
various sets of models. The components of GG~Tau were coeval when the models of 
BCAH98 with
$l_{mix}/H_p=1.0$ and 1.9 at $M/M_{\odot}\leq0.6$ and $M/M_{\odot}>0.6$
were combined with a temperature scale between those of dwarfs and giants.
\citet{whi99} also found that these models provided the best
agreement with the stellar masses estimated from the rotation
of circumstellar disks around GG~Tau Aa+Ab \citep{gui99}, 
GM~Aur \citep{dut98}, and DM~Tau \citep{gd98}.
\citet{luh99} updated the analysis of \citet{whi99} with revised 
spectral types and presented a temperature scale that produced coevality 
for GG~Tau with the models BCAH98. The models were also tested with
the empirical isochrone defined by the sequence of known members of IC~348,
the stellar parameters of the eclipsing double-lined spectroscopic binaries 
CM~Dra and YY~Gem, and the positions on the H-R diagram of brown dwarfs in 
the Pleiades. Both \citet{whi99} and \citet{luh99} concluded 
that the models of BCAH98 with $l_{mix}/H_p=1.9$ at $M/M_{\odot}>0.6$ 
agreed most closely with the various observational constraints.

Molecular line images of disks around several young stars have been obtained by 
\citet{gui99}, \citet{dut98}, \citet{gd98}, and \citet{sim00}.
From the disk rotation measured in these data, 
dynamical masses were estimated for a total of nine young systems.
\citet{sim00} placed these stars on the H-R diagram and compared
the masses implied by evolutionary models to those derived from the dynamical
measurements.  They found that the models of BCAH98,
\citet{pal99}, and \citet{sie00} were in reasonable agreement
with the observations, while the calculations of \citet{dm97}
were less consistent with the data.

Dynamical mass estimates from \citet{sim00} and from eclipsing 
binaries were used by \citet{pal01} to test the models from 
\citet{pal99}. They considered the main sequence binaries 
EW~Ori (1.2~$M_\odot$), HS~Aur (0.9~$M_\odot$), and YY~Gem (0.6~$M_\odot$)
and found agreement between the masses implied by their models and the
dynamical masses for first two systems but not the latter. 
Their comparison of the data and model predictions for YY~Gem has been
superseded by the thorough study of \citet{tor02}, who concluded that 
all available models implied physical parameters that differed significantly 
from the observed values. \citet{pal01} also examined young binaries, 
finding that the model predictions of \citet{pal99} 
agreed with the masses and mass ratios measured for four eclipsing and
three non-eclipsing young double-lined spectroscopic binaries at 1-6~$M_\odot$. 
Finally, the masses inferred from their models were within 8\% of the masses
derived from disk rotation \citep{sim00}, with the exception
of BP~Tau and UZ~Tau~E, whose dynamical masses have large uncertainties.

An additional pre-main-sequence eclipsing binary, RXJ0529.4+0041, has 
been discovered by \citet{cov00}. The components of this system 
have masses of $1.25\pm0.05$ and $0.91\pm0.05$~$M_\odot$ and spectral 
types of K1-K2 and K7-M0. When \citet{cov00} placed these stars 
on the H-R diagram, no set of models perfectly reproduced the dynamical 
mass estimates. The calculations of BCAH98 provided the best agreement.
\citet{dan00} have found that the inclusion of
magnetic fields in their models produces solar-mass tracks 
that are cooler and thus more closely match the data for RXJ0529.4+0041.

\citet{ste01} have measured masses for the components of the
pre-main-sequence binary NTT045251+3016 by combining radial velocity 
measurements and astrometry.
For the models of BCAH98, a mixing length of $l_{mix}/H_p=1.0$ produced the
closest agreement between the masses implied by the tracks on the H-R diagram 
and the dynamical masses of $1.45\pm0.19$ and $0.81\pm0.09$~$M_\odot$.
In contrast, the models with $l_{mix}/H_p=1.9$ better fit most other 
observations \citep{whi99,sim00}.
We note that \citet{ste01} used $V-H$ in their H-R diagram for 
NTT045251+3016 rather than effective temperature, which is probably not 
advisable given the shortcomings of the theoretical colors and magnitudes 
near $V$ for cool stars (e.g., \citet{del00}).

\subsection{Updated Tests of BCAH98 and CBAH00}

The previous tests of the evolutionary models generally favor the 
calculations of BCAH98, particularly below a solar mass.
In this section, the best available constraints at young ages are 
compiled from previous studies and from this work on IC~348 and are 
applied to these models in a consistent fashion.

In this discussion, we consider the models of BCAH98
with $l_{mix}/H_p=1.0$ and 1.9 for $0.1<M/M_\odot\leq0.6$ and
$0.6<M/M_\odot\leq1.4$ and the models of CBAH00 with
$l_{mix}/H_p=1.0$ for $0.001\leq M/M_\odot\leq0.1$. 
The calculations of BCAH98 and CBAH00 exclude and include 
dust, respectively, and produce similar mass tracks and isochrones for 
the effective temperatures of 2000-3000~K where they overlap \citep{bar02}.
BCAH98 found that a convection mixing length of 1.9
was required to reproduce data for the Sun. Most of the
observational constraints for young stars from the previous section 
were best matched with this mixing length as well. For masses below
0.6~$M_\odot$, BCAH98 computed models only for
$l_{mix}/H_p=1.0$ because the results were not sensitive to the choice of 
mixing length for low masses and older ages. However, in a recent analysis of
the uncertainties in their model predictions, \citet{bar02}
derived tracks at low masses that changed significantly with mixing length
for log~$g\lesssim4$. For instance, at an age of 1~Myr and masses of
0.01-0.2~$M_\odot$, the mass tracks for $l_{mix}/H_p=2.0$ were 100-200~K 
warmer than those with $l_{mix}/H_p=1.0$. \citet{bar02} computed
only a restricted grid of models at $l_{mix}/H_p=2.0$ to demonstrate the
effect of changes in the mixing length. Thus, low-mass models covering the full 
range of masses and ages are available only for $l_{mix}/H_p=1.0$ at this time.
Compared to the convection mixing length, changes in the initial radius and 
the initial deuterium abundance have less effect on the mass tracks and
isochrones at ages of $\gtrsim1$~Myr \citep{bar02}.

For testing the models of BCAH98 and CBAH00, we consider all 
pre-main-sequence stars that have reasonably accurate dynamical mass estimates 
and are below 1.5~$M_\odot$, which include five components of three 
spectroscopic binaries (RXJ0529.4+0041 A and B, NTT045251+3016 A and B,
EK~Cep~B) and five systems from \citet{sim00} 
(DL~Tau, DM~Tau, LkCa~15, GM~Aur, GG~Tau~Aa+Ab).
We omit the double-lined spectroscopic binaries for which mass ratios alone are 
available \citep{cov01,pra02}.
We compile luminosities and temperatures for this sample of stars 
in the manner described in \S~\ref{sec:append3}.
These measurements and the models of BCAH98 and CBAH00 are plotted on 
the H-R diagram in Figure~\ref{fig:hr1}. Before testing the masses inferred
from the models with the dynamical masses, we first compare the latter
measurements to each other. The relative positions of most of the stars in 
Figure~\ref{fig:hr1} are consistent with their relative dynamical masses.
The clear exceptions are the components of NTT045251+3016, which have
dynamical masses that are anomalously high for their positions in the H-R
diagram relative to the other stars. This discrepancy is reflected in the 
conclusion by \citet{ste01} that the data for NTT045251+3016 
were best fit by the BCAH98 models with $l_{mix}/H_p=1.0$,
whereas the studies of the other stars with dynamical masses favored 
$l_{mix}/H_p=1.9$. If we exclude NTT045251+3016, then the models of
BCAH98 and CBAH00 are in fairly good agreement with the data. For most
of the stars, the masses implied by the models and the dynamical masses
agree within the uncertainties. An exception is the primary of RXJ0529.4+0041,
which has dynamical and model masses of $1.25\pm0.05$ and 
$\gtrsim1.4$~$M_\odot$, respectively. The data for RXJ0529.4+0041~A
and the other star above a solar mass in Figure~\ref{fig:hr1}, EK~Cep~B, 
are better matched by the models of \citet{pal99}.

As described in the previous section, \citet{whi99} and \citet{luh99}
tested evolutionary models by posing the following question:
For a given set of models, is there a reasonable temperature scale that will
make an empirical isochrone parallel to the model isochrones?
The models of BCAH98 passed this test for the empirical
isochrones defined by the GG~Tau quadruple system and the IC~348 cluster.
The limit in spectral type of this test was M7.5 because 
this was the latest type present in GG~Tau and because the list of known
members of IC~348 in \citet{luh99} was well-populated down to only M6. 
We have extended this test to M9 by using the new samples of low-mass members 
of Taurus and IC~348 from \citet{bri02}, \citet{luh03}, and this work.
In \S~\ref{sec:teff}, we were able to adjust the temperature scale from 
\citet{luh99} at M8 and M9 so that the sequences for Taurus and IC~348 are
roughly parallel to the model isochrones for the full range of observed 
spectral types. This is illustrated in Figure~\ref{fig:hr2}, 
where we plot H-R diagrams for the extinction-limited samples that define the 
IMFs computed for Taurus \citep{bri02,luh03} and IC~348 (\S~\ref{sec:imf348}).

Overall, the best available observational constraints for the masses and
ages of young stars below a solar mass tend to support the validity of the 
models of BCAH98 and CBAH00. 
The data favor the models with a convection mixing length of $l_{mix}/H_p=1.9$. 
However, this choice of mixing length is not available for models
at $M/M_\odot<0.6$. As a result, the models at low masses, which use 
$l_{mix}/H_p=1.0$, may be less accurate than at higher masses. We cannot
test the accuracy of the low-mass tracks since dynamical masses have not 
been measured in this range.
Any errors in the low-mass models should be at least partially compensated by 
our use of a temperature scale that is designed to produce populations
that are parallel to the model isochrones. For instance, 
it is possible that the true temperature scale for young objects is 
that of dwarfs, and the deviation of our conversion from the dwarf scale 
(Figure~\ref{fig:scale}) is a reflection of an error in the models
(e.g., use of $l_{mix}/H_p=1.0$).
Further tests of the models at young ages and low masses will require 
the measurement of dynamical masses and the modeling of both additional mixing 
lengths and a birthline for brown dwarfs (e.g., \citet{sta83}).

\section{Data for Stars with Dynamical Mass Estimates}
\label{sec:append3}

\begin{figure}
\epsscale{0.8}
\plotone{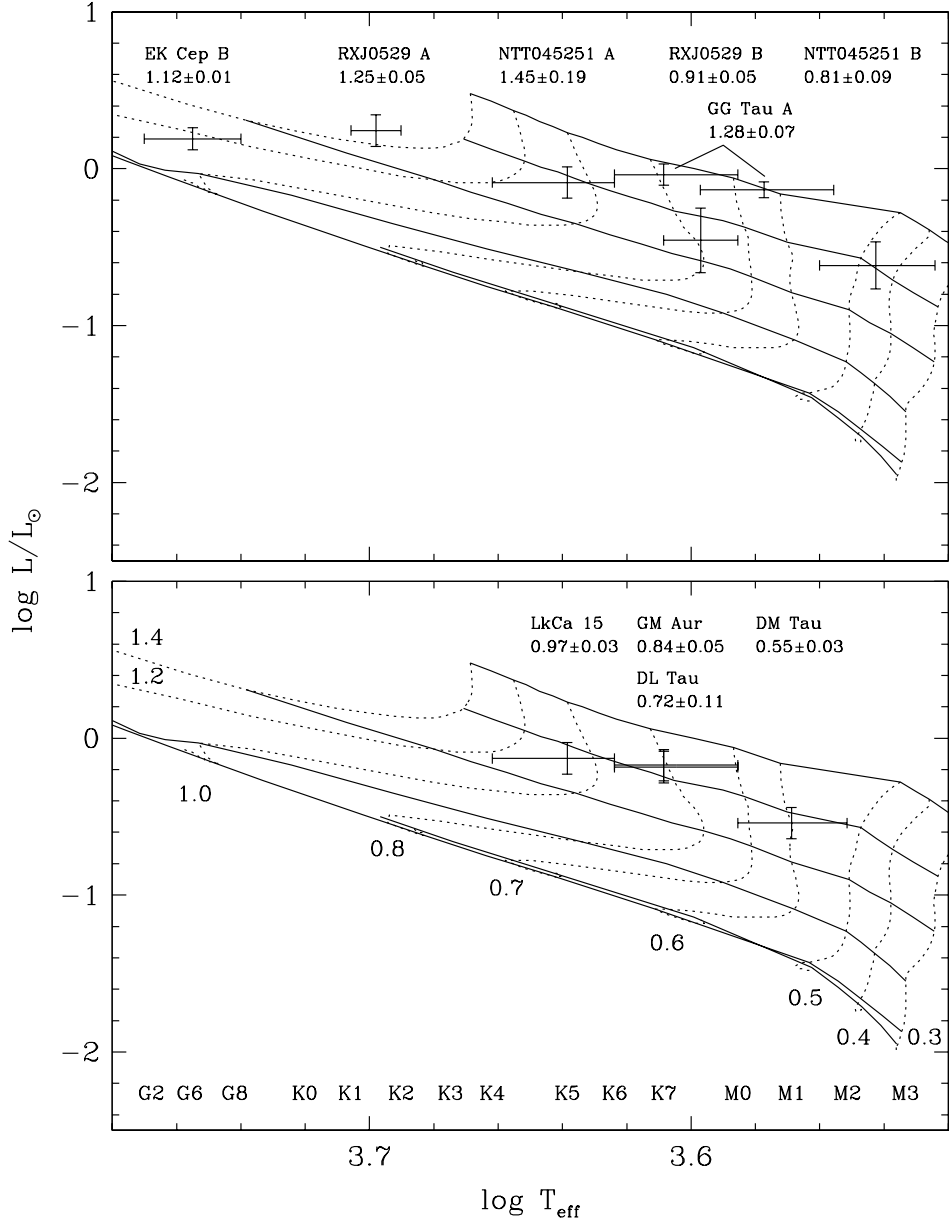}
\caption{
Young stars that have dynamical mass estimates are shown with the evolutionary 
models of BCAH98 ($M/M_{\odot}>0.1$) and CBAH00 ($M/M_{\odot}\leq0.1$)
with $l_{mix}/H_p=1.0$ ($M/M_\odot\leq0.6$) and 1.9 ($M/M_\odot>0.6$).
The horizontal solid lines are isochrones representing ages of 1, 3, 10, 30,
and 100~Myr and the main sequence, from top to bottom. For clarity, 
these sources are divided into single and binary stars ({\it lower and upper}).
Mass estimates in solar masses are indicated, which are from 
\citet{sim00} (LkCa~15, GM~Aur, DL~Tau, DM~Tau,
GG~Tau~A), \citet{ste01} (NTT045251+3016), \citet{cov00}
(RXJ0529.4+0041), and \citet{pop87} (EK~Cep). 
The mass listed for GG~Tau~A is the combined mass of Aa and Ab, which are 
plotted individually.
}
\label{fig:hr1}
\end{figure}

We describe the derivation of the effective temperatures and luminosities 
that are used in placing RXJ0529.4+0041 A and B, NTT045251+3016 A and B, 
EK~Cep~B, DL~Tau, DM~Tau, LkCa~15, GM~Aur, and GG~Tau~Aa+Ab on the H-R 
diagram in Figure~\ref{fig:hr1}. Spectral types are converted to
temperatures with the scale in \S~\ref{sec:teff}.
For the two components of RXJ0529.4+0041, we adopt spectral types 
of K1-K2 and K7-M0 and luminosities of $1.75\pm0.15$ and 
$0.35\pm0.15$~$L_\odot$ from \citet{cov00}. 
Those authors identified a possible third component of the system and placed
it on the H-R diagram, but did not provide photometry or a luminosity estimate
for it. The uncertainties listed by \citet{cov00} for the primary's 
luminosity are much smaller than the values plotted on their H-R diagram.
The mistake is probably in $\sigma_L=\pm0.15$~$L_\odot$ 
since the quoted measurement errors for the radius alone ($1.7\pm0.2$~$R_\odot$)
would correspond to $\sigma_L=\pm0.4$~$L_\odot$. We adopt the latter value for
the uncertainty in the luminosity. For the primary of NTT045251+3016,
we use the spectral type of K5 from \citet{ste01}
and adopt an uncertainty of $\pm1$~subclass. From the intrinsic color of
$V-H=4.04\pm0.33$ estimated for the secondary by \citet{ste01},
we infer a spectral type of M2.5$\pm1$. Those authors listed uncertainties in 
log~$L$ of $\pm0.053$ and $\pm0.086$, which appear to be unrealistically small
given that the distance uncertainties alone correspond to $\pm0.048$
in log~$L$. In addition, \citet{wal88} reports $H=8.46$ for the system
while 2MASS measures $H=8.32$. Therefore, we adopt uncertainties of $\pm0.1$
and $\pm0.15$ in log~$L$ for the primary and secondary. 
We take $A_V=0.15\pm0.09$ and $d=145\pm8$~pc from \citet{ste01}.
Using their $H$-band brightness ratio of $0.4\pm0.1$ and a total system 
magnitude of $H=8.32$ from 2MASS, we calculate $H=8.69$ and 9.68 for the 
individual components. Luminosities are derived by combining these $H$-band 
magnitudes with the distance and the appropriate bolometric corrections
(\S~\ref{sec:teff}). 
For the secondary in the spectroscopic binary EK~Cep, we adopt 
log~$L=0.19\pm0.07$ and log~$T_{\rm eff}=3.755\pm0.015$ from \citet{pop87}.
Temperatures and luminosities for DL~Tau, DM~Tau, LkCa~15, and GM~Aur 
are computed with the prescription described by \citet{bri02}
for members of Taurus. We use the same distance of 140~pc that was
adopted in the dynamical mass estimates by \citet{sim00}.
We note that \citet{kh95} mistakenly listed $R-I=1.53$ for GM~Aur 
instead of the correct value of $R-I=0.72$ (1.53 is the measurement of $V-I$).
Uncertainties of $\pm1$~subclass and $\pm0.1$ are adopted for the spectral
types and for log~$L$, respectively.
For GG~Tau~Aa and Ab, we adopt extinctions, spectral types, luminosity 
uncertainties, and photometry from \citet{whi99} and compute
luminosities by combining the $J$-band photometry with the extinctions,
bolometric corrections, and a distance of 140~pc.

\clearpage

\begin{deluxetable}{llllll}
\tabletypesize{\scriptsize}
\tablewidth{0pt}
\tablecaption{Observing Log \label{tab:log}}
\tablehead{
\colhead{Date} &
\colhead{Telescope + Instrument} &
\colhead{Grating} &
\colhead{Resolution} &
\colhead{ID} \\
\colhead{} &
\colhead{} &
\colhead{l~mm$^{-1}$} &
\colhead{\AA} &
\colhead{}
}
\startdata
1998 Aug 7 & Keck + LRIS & 150 & 16 & 384,547,548,550,631   \\
1998 Nov 24 & Keck + LRIS & 1200 & 2.5 & 435 \\
1998 Dec 23, 26 & KPNO 4 m + RCSP & 158 & 14 & 31,43,48,53,55,61,69,79,81,82,84,89,92,93,105,116,119 \\
\nodata & \nodata & \nodata & \nodata & 
123,129,146,174,177,186,188,210,212,236,246,281,282,311   \\
\nodata & \nodata & \nodata & \nodata &
317,339,369,462,829 \\
1999 Dec 23 & Keck + LRIS & 150 & 16 & 301,382,391   \\
2000 Dec 3 & Steward 2.3 m + B\&C & 400 & 7 & 56,63,66,96,97,98,106,156,182,195,198,206,228,253,254,255   \\
2000 Dec 7 & MMT + FSpec & 75 & 27 & 27,43,84,89,143,219   \\
\enddata
\end{deluxetable}

\begin{deluxetable}{llllllllllllllllll}
\setlength{\tabcolsep}{0.04in}
\tabletypesize{\scriptsize}
\rotate
\tablewidth{0pt}
\tablecaption{Data for Members of IC 348 \label{tab:mem}}
\tablehead{
\colhead{ID} &
\colhead{$\alpha$(J2000)\tablenotemark{a}} & 
\colhead{$\delta$(J2000)\tablenotemark{a}} &
\colhead{Spectral Type/$W_{\lambda}$(H$\alpha$)\tablenotemark{b}} &
\colhead{Ref} &
\colhead{Adopt} &
\colhead{Member\tablenotemark{c}} &
\colhead{$T_{\rm eff}$\tablenotemark{d}} &
\colhead{$A_J$} & \colhead{$L_{\rm bol}$} &
\colhead{$R-I$\tablenotemark{e}} & \colhead{$I$\tablenotemark{f}} &
\colhead{$I-Z$\tablenotemark{g}} & 
\colhead{$J-H$} & \colhead{$H-K_s$} & \colhead{$K_s$} &
\colhead{Ref} &
\colhead{IMF?} \\
\colhead{} &
\colhead{h m s} &
\colhead{$\arcdeg$ $\arcmin$ $\arcsec$} &
\colhead{} &
\colhead{} &
\colhead{} &
\colhead{-ship} &
\colhead{} &
\colhead{} &
\colhead{} &
\colhead{} &
\colhead{} &
\colhead{} &
\colhead{} &
\colhead{} &
\colhead{} &
\colhead{} &
\colhead{}
}
\startdata
       1$\tablenotemark{h}$ &   03 44 34.20 &    32 09 46.3 &    B5V,B3-B4,$<$F8(IR) & 1,2,2 &         B5 &      $A_V$,pm &  15400 &    0.54 &     605 &      \nodata &       \nodata &       \nodata &     0.15 &     0.12 &     6.51 &  3 &    yes  \\
       2 &   03 44 35.36 &    32 10 04.6 &           A2,$<$F8(IR) & 1,2 &         A2 &      $A_V$,pm &   8970 &    1.08 &     137 &      \nodata &       \nodata &       \nodata &     0.42 &     0.29 &     7.25 &  3 &    yes  \\
       3 &   03 44 50.65 &    32 19 06.8 &           A0,$<$F8(IR) & 4,2 &         A0 &         $A_V$ &   9520 &    1.26 &     135 &      \nodata &       \nodata &       \nodata &     0.48 &     0.24 &     7.66 &  3 &     no  \\
       4 &   03 44 31.19 &    32 06 22.1 &         F0m:,$<$F8(IR) & 1,2 &         F0 &      $A_V$,pm &   7200 &    0.50 &      34 &      \nodata &       \nodata &       \nodata &     0.31 &     0.16 &     7.86 &  3 &    yes  \\
       5 &   03 44 26.03 &    32 04 30.4 & K0/23,F8-G8/2.1,?(IR) & 5,2,2 &         G8 &   $A_V$,ex,e &   5520 &    1.50 &     9.9 &    1.35 &    12.44 &     0.72 &     1.21 &     0.73 &     8.14 &  3 &     no  \\
       6$\tablenotemark{h}$ &   03 44 36.94 &    32 06 45.4 &   F8,F8-G8,G3-G7(IR) & 4,2,2 &         G3 &      $A_V$,pm &   5830 &    1.10 &      17 &      \nodata &       \nodata &       \nodata &     0.72 &     0.29 &     8.19 &  3 &    yes  \\
       7 &   03 44 08.48 &    32 07 16.5 &          A0V,$<$F8(IR) & 1,2 &         A0 &      $A_V$,pm &   9520 &    0.26 &      35 &      \nodata &       \nodata &       \nodata &     0.10 &     0.12 &     8.60 &  3 &    yes  \\
       8 &   03 44 09.15 &    32 07 09.3 &          A2V,$<$F8(IR) & 1,2 &         A2 &      $A_V$,pm &   8970 &    0.27 &      28 &      \nodata &       \nodata &       \nodata &     0.11 &     0.14 &     8.62 &  3 &    yes  \\
\enddata
\end{deluxetable}

\begin{deluxetable}{llllllllllll}
\tabletypesize{\scriptsize}
\tablewidth{0pt}
\tablecaption{Foreground Stars \label{tab:fore}}
\tablehead{
\colhead{ID} &
\colhead{$\alpha$(J2000)\tablenotemark{a}} & 
\colhead{$\delta$(J2000)\tablenotemark{a}} &
\colhead{Spectral Type} &
\colhead{Ref} &
\colhead{Foreground\tablenotemark{b}} &
\colhead{$R-I$\tablenotemark{c}} & \colhead{$I$\tablenotemark{d}} &
\colhead{$I-Z$\tablenotemark{d}} & 
\colhead{$J-H$\tablenotemark{e}} & \colhead{$H-K_s$\tablenotemark{e}} 
& \colhead{$K_s$\tablenotemark{e}} \\
\colhead{} &
\colhead{h m s} &
\colhead{$\arcdeg$ $\arcmin$ $\arcsec$} &
\colhead{} &
\colhead{} &
\colhead{Evidence} &
\colhead{} &
\colhead{} &
\colhead{} &
\colhead{} &
\colhead{} &
\colhead{} 
}
\startdata
      18 &   03 43 58.04 &    32 09 47.1 &                   F8 &    1 &         pm &      \nodata &       \nodata &       \nodata &     0.28 &     0.13 &     9.36  \\
      28 &   03 44 21.06 &    32 07 38.7 &                   F8 &    1 &         pm &      \nodata &       \nodata &       \nodata &     0.30 &     0.11 &     9.62  \\
      54 &   03 44 49.59 &    32 16 45.7 &                K1-K2 &    2 &      Li,$A_V$ &      \nodata &    12.04 &     0.32 &     0.50 &     0.16 &    10.50  \\
      57 &   03 44 48.51 &    32 15 28.1 &                K3-K4 &    2 &         pm &      \nodata &    12.45 &     0.36 &     0.68 &     0.15 &    10.56  \\
      77 &   03 44 43.44 &    32 08 17.4 &         M1,K5-M0(IR) & 3,2 &         pm &    0.96 &    12.90 &     0.37 &     0.62 &     0.23 &    10.85  \\
     107 &   03 43 55.81 &    32 19 44.3 &         M1,K6-M1(IR) & 2,2 &         pm &      \nodata &    13.33 &     0.45 &     0.68 &     0.26 &    11.21  \\
     121 &   03 44 25.95 &    32 16 30.7 &            K4-K7(IR) &    2 &         pm &      \nodata &    13.39 &     0.37 &     0.80 &     0.12 &    11.36  \\
     134 &   03 43 59.88 &    32 11 16.2 &       M3.5,M2-M4(IR) & 2,2 &         pm &      \nodata &    13.56 &     0.49 &     0.64 &     0.20 &    11.34  \\
\enddata
\tablenotetext{a}{From the 2MASS Point Source Catalog for 18 and 28. 
All other coordinates are from the $I$-band images in this work.}
\tablenotetext{b}{Foreground nature is indicated by proper motion 
measurements (``pm"; \citet{bla52,sch99}) or a combination 
of no Li absorption (``Li"), $A_V\sim0$ (``$A_V$"), and spectral features 
such as strong Na~I and K~I (``NaK"; \citet{luh98b}; this work).}
\tablenotetext{c}{From \citet{her98} for 77 and from \citet{luh99}
for the remaining objects.}
\tablenotetext{d}{From this work.}
\tablenotetext{e}{From 2MASS for 18, 28, 54, 57, 77, 189, 320, 400, and 1788 
and from \citet{mue03} for the remaining objects.}
\tablerefs{(1) \citet{fre56}; (2) \citet{luh98b}; (3) \citet{her98};
(4) Measurements described by \citet{luh99} and presented here for the first
time; (5) this work.}
\end{deluxetable}

\begin{deluxetable}{lllllllllll}
\tabletypesize{\scriptsize}
\tablewidth{0pt}
\tablecaption{Background Stars \label{tab:back}}
\tablehead{
\colhead{ID} &
\colhead{$\alpha$(J2000)\tablenotemark{a}} & 
\colhead{$\delta$(J2000)\tablenotemark{a}} &
\colhead{Spectral Type} &
\colhead{Ref} &
\colhead{$R-I$\tablenotemark{b}} & \colhead{$I$\tablenotemark{c}} &
\colhead{$I-Z$\tablenotemark{d}} & 
\colhead{$J-H$\tablenotemark{e}} & \colhead{$H-K_s$\tablenotemark{e}} 
& \colhead{$K_s$\tablenotemark{e}} \\
\colhead{} &
\colhead{h m s} &
\colhead{$\arcdeg$ $\arcmin$ $\arcsec$} &
\colhead{} &
\colhead{} &
\colhead{} &
\colhead{} &
\colhead{} &
\colhead{} &
\colhead{} &
\colhead{} 
}
\startdata
      14 &   03 44 54.95 &    32 12 11.3 &  K0-1III,K1-2III(IR) &    1,1 &      \nodata &    12.77 &     0.77 &     1.04 &     0.43 &     8.93  \\
      27 &   03 43 36.16 &    32 13 32.9 &      F7-G6,giant(IR) &    1,2 &      \nodata &    12.57 &     0.54 &     0.83 &     0.29 &     9.68  \\
      34 &   03 44 15.48 &    32 02 35.1 &            G8III(IR) &        1 &    2.28 &    16.91 &     1.18 &     1.98 &     0.98 &     9.89  \\
      43 &   03 45 03.19 &    32 04 05.7 &         giant(op,IR) &   3,2 &      \nodata &    14.64 &     0.83 &     1.21 &     0.45 &    10.29  \\
      73 &   03 45 09.64 &    32 16 03.6 &                giant &        2 &      \nodata &    13.73 &     0.55 &     0.83 &     0.28 &    10.81  \\
      80 &   03 44 27.88 &    32 10 52.2 &            K1III(IR) &        1 &    2.60 &    18.52 &     1.35 &     2.19 &     1.03 &    10.87  \\
      84 &   03 44 42.14 &    32 02 57.8 &         giant(op,IR) &   3,2 &      \nodata &    15.04 &     0.74 &     1.19 &     0.39 &    11.01  \\
      89 &   03 44 57.58 &    32 12 24.4 &         giant(op,IR) &   3,2 &      \nodata &    14.40 &     0.71 &     0.99 &     0.28 &    11.04  \\
\enddata
\tablenotetext{a}{From the 2MASS Point Source Catalog
for 206 and 370 and from \citet{luh99} for 842.
All other coordinates are from the $I$-band images in this work.}
\tablenotetext{b}{From \citet{her98} for 106 and 206 and from \citet{luh99}
for the remaining objects.}
\tablenotetext{c}{From \citet{her98} for 206, from \citet{luh99} for 370 and 
842, and from this work for the remaining objects.}
\tablenotetext{d}{From this work.}
\tablenotetext{e}{From 2MASS for 14, 27, 34, 43, 73, 80, 131, 1932, 1941, 
10226, 10252, and 10338 and from \citet{mue03} for the remaining objects.}
\tablerefs{(1) \citet{luh98b}; (2) this work;
(3) Measurements described by \citet{luh99} and presented here for the first
time.}
\end{deluxetable}

\begin{deluxetable}{llllllllllll}
\tabletypesize{\scriptsize}
\tablewidth{0pt}
\tablecaption{Sources with Uncertain Membership Status \label{tab:unc}}
\tablehead{
\colhead{ID} &
\colhead{$\alpha$(J2000)\tablenotemark{a}} & 
\colhead{$\delta$(J2000)\tablenotemark{a}} &
\colhead{Spectral Type} &
\colhead{Ref} &
\colhead{Status\tablenotemark{b}} &
\colhead{$R-I$\tablenotemark{c}} & \colhead{$I$\tablenotemark{d}} &
\colhead{$I-Z$\tablenotemark{d}} & 
\colhead{$J-H$\tablenotemark{e}} & \colhead{$H-K_s$\tablenotemark{e}} 
& \colhead{$K_s$\tablenotemark{e}} \\
\colhead{} &
\colhead{h m s} &
\colhead{$\arcdeg$ $\arcmin$ $\arcsec$} &
\colhead{} &
\colhead{} &
\colhead{} &
\colhead{} &
\colhead{} &
\colhead{} &
\colhead{} &
\colhead{} &
\colhead{} 
}
\startdata
      81 &   03 44 50.00 &    32 03 45.7 &         F2-G2 or giant &  1 &         m or b &      \nodata &    15.08 &     0.81 &     1.04 &     0.46 &    10.97  \\
     197 &   03 44 29.11 &    32 07 51.0 &            G6-K1(IR) &   2 &         m or b &    2.02 &    17.36 &     1.00 &     1.37 &     0.61 &    12.15  \\
    1927 &   03 45 07.74 &    32 00 27.6 &                K3-K4 &   3 &         m or f &      \nodata &    11.71 &     0.36 &     0.63 &     0.17 &     9.67  \\
\enddata
\tablenotetext{a}{From the $I$-band images in this work.}
\tablenotetext{b}{m=member, b=background, f=foreground.}
\tablenotetext{c}{From \citet{luh99}.}
\tablenotetext{d}{From this work.}
\tablenotetext{e}{From 2MASS for 81 and 1927 and from \citet{mue03} for 197.}
\tablerefs{(1) Measurements described by \citet{luh99} and presented here for
the first time; (2) \citet{luh98b}; (3) this work.}
\end{deluxetable}

\begin{deluxetable}{lllll}
\tabletypesize{\scriptsize}
\tablewidth{0pt}
\tablecaption{Recent Identifications for Sources in IC 348 \label{tab:names}}
\tablehead{
\colhead{ID\tablenotemark{a}} &
\colhead{\citet{her98}} &
\colhead{\citet{naj00}} &
\colhead{\citet{pre02}} &
\colhead{2MASS}
}
\startdata
   1 &    \nodata &      \nodata &    \nodata & J03443419+3209462   \\
   2 &    \nodata &      \nodata &   124 & J03443536+3210045   \\
   3 &    \nodata &      \nodata &   187 & J03445064+3219067   \\
   4 &   254 &      \nodata &    98 & J03443118+3206220   \\
   5 &   114 &      \nodata &    73 & J03442602+3204304   \\
   6 &   166 &      \nodata &   133 & J03443694+3206453   \\
   7 &    \nodata &      \nodata &    \nodata & J03440847+3207165   \\
   8 &    \nodata &      \nodata &    \nodata & J03440915+3207093   \\
\enddata
\tablenotetext{a}{Identifications from \citet{luh98b}, \citet{luh99}, and
this work.}
\end{deluxetable}

\begin{deluxetable}{lll}
\tablewidth{0pt}
\tablecaption{Adopted Intrinsic Colors \label{tab:iz}}
\tablehead{
\colhead{Spectral Type} &
\colhead{$R-I$} & \colhead{$I-Z$} 
}
\startdata
K0 & 0.39 & 0.23 \\
K1 & 0.43 & 0.25 \\
K2 & 0.47 & 0.27 \\
K3 & 0.50 & 0.29 \\
K4 & 0.50 & 0.31 \\
K5 & 0.63 & 0.33 \\
K6 & 0.70 & 0.35 \\
K7 & 0.76 & 0.37 \\
M0 & 0.75 & 0.40 \\
M1 & 0.92 & 0.43 \\
M2 & 1.00 & 0.46 \\
M3 & 1.10 & 0.49 \\
M4 & 1.25 & 0.55 \\
M5 & 1.65 & 0.67 \\
M6 & 2.10 & 0.85 \\
M7 & 2.30 & 0.98 \\
M8 & 2.50 & 1.10 \\
M9 & 2.50 & 1.21 \\
\enddata
\end{deluxetable}

\begin{deluxetable}{ll}
\tablewidth{0pt}
\tablecaption{Adopted Temperature Scale \label{tab:scale}}
\tablehead{\colhead{Spectral Type} & \colhead{$T_{\rm eff}$(K)}}
\startdata
M1 & 3705 \\
M2 & 3560 \\
M3 & 3415 \\
M4 & 3270 \\
M5 & 3125 \\
M6 & 2990 \\
M7 & 2880 \\
M8 & 2710 \\
M9 & 2400 \\
\enddata
\end{deluxetable}

\begin{deluxetable}{llll}
\tablewidth{0pt}
\tablecaption{Frequencies of Brown Dwarfs and High-Mass Stars \label{tab:ratios}}
\tablehead{\colhead{Population} & 
\colhead{${\mathcal R}_{1}=\frac{N(0.02-0.08)}{N(0.08-10)}$} & 
\colhead{${\mathcal R}_{2}=\frac{N(1-10)}{N(0.15-1)}$} & 
\colhead{Reference}} 
\startdata
Taurus & $0.14\pm0.04$ & $0.08\pm0.04$ & 1,2 \\
IC~348 & $0.12\pm0.03$ & $0.18\pm0.04$ & 3 \\
Trapezium & $0.26\pm0.04$ & $0.21\pm0.04$ & 4 \\
Pleiades & \nodata & $0.27\pm0.02$ & 5 \\
M35 & \nodata & $0.22\pm0.01$ & 6 \\
\enddata
\tablerefs{(1) \citet{bri02}; (2) \citet{luh03}; (3) this work;
(4) \citet{luh00b}; (5) \citet{bou98}; (6) \citet{bar01}.} 
\end{deluxetable}

\clearpage

\clearpage

\clearpage

\clearpage

\clearpage

\clearpage

\clearpage

\clearpage

\clearpage

\clearpage

\clearpage

\clearpage

\clearpage

\clearpage

\clearpage


\newpage

\begin{thebibliography}{}

\bibitem[Adams \& Fatuzzo(1996)]{af96}
Adams, F. C., \& Fatuzzo, M. 1996, \apj, 464, 256

\bibitem[Allard et al.(2001)]{all01}
Allard, F., Hauschildt, P. H., Alexander, D. R., Tamanai, A., 
\& Schweitzer, A. 2001, \apj, 556, 357

\bibitem[Allen \& Strom(1995)]{as95}
Allen, L. E., \& Strom, K. M. 1995, \aj, 109, 1379

\bibitem[Bachiller, Guilloteau, \& Kahane(1987)]{bac87}
Bachiller, R., Guilloteau, S., \& Kahane, C. 1987, \aap, 173, 324

\bibitem[Baraffe et al.(1998)]{bar98}
Baraffe, I., Chabrier, G., Allard, F., \& Hauschildt, P. H. 1998, \aap, 337, 403
(BCAH98)

\bibitem[Baraffe et al.(2002)]{bar02}
Baraffe, I., Chabrier, G., Allard, F., \& Hauschildt, P. H. 2002, \aap, 382, 563

\bibitem[Barrado y Navascu\'es et al.(2001)]{bar01}
Barrado y Navascu\'es, D., Stauffer, J. R., Brice\~no, C., 
Patten, B., Hambly, N. \& Adams, J. 2001, \apjs, 134, 103

\bibitem[Bessell \& Brett(1988)]{bb88}
Bessell, M. S., \& Brett, J. M. 1988, \pasp, 100, 1134

\bibitem[Blaauw(1952)]{bla52}
Blaauw, A. 1952, B. A. N., 11, 412

\bibitem[Bouvier et al.(1998)]{bou98}
Bouvier, J., Stauffer, J. R., Mart{\'\i}n, E. L., Barrado y 
Navascu\'{e}s, D., Wallace, B., \& Bejar, V. J. S. 1998, \aap, 336, 490

\bibitem[Brice\~no et al.(2002)]{bri02}
Brice\~{n}o, C., Luhman, K. L., Hartmann, L., Stauffer, J. R., \& Kirkpatrick,
J. D. 2002, \apj, 580, 317

\bibitem[Brown et al.(1989)]{bro89}
Brown, J. A., Sneden, C., Lambert, D. L., \& Dutchover, E. 1989, \apjs, 71, 293

\bibitem[Burgasser et al.(2002)]{bur02}
Burgasser, A., et al. 2002, \apj, 564, 421

\bibitem[Burrows et al.(1997)]{bur97}
Burrows, A., et al. 1997, \apj, 491, 856

\bibitem[Carpenter(2001)]{car01}
Carpenter, J. M. 2001, \aj, 121, 2851

\bibitem[Carpenter(2002)]{car02}
Carpenter, J. M. 2002, \aj, 124, 1593

\bibitem[Chabrier et al.(2000)]{cha00}
Chabrier, G., Baraffe, I. Allard, F., \& Hauschildt, P. H. 2000,
\apj, 542, 464 (CBAH00)

\bibitem[Covino et al.(2000)]{cov00}
Covino, E., et al. 2000, \aap, 361, L49

\bibitem[Covino et al.(2001)]{cov01}
Covino, E., Melo, C., Alcal\'a, J. M., Torres, G., Fern\'andez, M.,
Frasca, A., \& Paladino, R. 2001, \aap, 375, 130

\bibitem[D'Antona \& Mazzitelli(1997)]{dm97}
D'Antona, F., \& Mazzitelli, I. 1997, in Cool Stars in
Clusters and Associations, eds. G. Micela, R. Pallavicini \& S. Sciortino,
Mem. Soc. Astron. Italiana, 68, 807

\bibitem[D'Antona et al.(2000)]{dan00}
D'Antona, F., Ventura, P., \& Mazzitelli, I. 2000, \apj, 543, L77

\bibitem[Delfosse et al.(2000)]{del00}
Delfosse, X., Forveille, T., S\'egransan, D., Beuzit, J.-L., 
Udry, S., Perrier, C., \& Mayor, M. 2000, \aap, 364, 217

\bibitem[Duch\^ene et al.(1999)]{duc99}
Duch\^ene, G., Bouvier, J., \& Simon, T. 1999, 343, 831

\bibitem[Dutrey et al.(1998)]{dut98}
Dutrey, A., Guilloteau, S., Prato, L., Simon, M., Duvert, G.,
Schuster, K., \& Menard, F. 1998, \aap, 338, 63

\bibitem[Fabricant et al.(1998)]{fab98}
Fabricant, D., Cheimets, P., Caldwell, N., \& Geary, J. 1998, \pasp, 110, 79

\bibitem[Fredrick(1956)]{fre56}
Fredrick, L. W. 1956, \aj, 61, 437

\bibitem[Gingrich(1922)]{gin22}
Gingrich, C. H. 1922, \apj, 56, 139

\bibitem[Guilloteau \& Dutrey(1998)]{gd98}
Guilloteau, S., \& Dutrey, A. 1998, \aap, 339, 467

\bibitem[Guilloteau et al.(1999)]{gui99}
Guilloteau, S., Dutrey, A., \& Simon M. 1999, \aap, 348, 570

\bibitem[Haisch et al.(2000)]{hai00}
Haisch, K. E., Lada, E. A., \& Lada, C. J. 2000, \aj, 120, 1396

\bibitem[Haisch et al.(2001)]{hai01}
Haisch, K. E., Lada, E. A., \& Lada, C. J. 2001, \aj, 121, 2065

\bibitem[Harris et al.(1954)]{har54}
Harris, D. L., Morgan, W. W., \& Roman, N. G. 1954, \apj, 119, 622

\bibitem[Hartigan et al.(1994)]{har94}
Hartigan, P., Strom, K. M., \& Strom, S. E. 1994, \apj, 427, 961

\bibitem[Hartmann(2001)]{har01}
Hartmann, L. 2001, \aj, 121, 1030

\bibitem[Henry et al.(1994)]{hen94}
Henry, T. J., Kirkpatrick, J. D., \& Simons, D. A. 1994, \aj, 108, 1437

\bibitem[Herbig(1954)]{her54}
Herbig, G. H. 1954, \pasp, 66, 19

\bibitem[Herbig(1998)]{her98}
Herbig, G. H. 1998, \apj, 497, 736

\bibitem[Herbst et al.(2000)]{her00}
Herbst, W., Maley, J. A., \& Williams, E. C. 2000, \aj, 120, 349

\bibitem[Hillenbrand(1997)]{hil97}
Hillenbrand, L. A. 1997, \aj, 113, 1733

\bibitem[Hillenbrand \& Carpenter(2000)]{hc00}
Hillenbrand, L. A., \& Carpenter, J. M. 2000, \apj, 540, 236

\bibitem[H\"{u}ensch et al.(1996)]{hue96}
H\"{u}ensch, M., Schmitt, J. H. M. M., Schroeder, K.-P., \& Reimers, D. 1996, 
\aap, 310, 801

\bibitem[Kalas \& Jewitt(1997)]{kal97}
Kalas, P., \& Jewitt, D. 1997, Nature, 386, 52

\bibitem[Kenyon \& Hartmann(1990)]{kh90}
Kenyon, S. J., \& Hartmann, L. 1990, \apj, 349, 197

\bibitem[Kenyon \& Hartmann(1995)]{kh95}
Kenyon, S. J., \& Hartmann, L. 1995, \apjs, 101, 117

\bibitem[Kirkpatrick et al.(1997)]{kir97}
Kirkpatrick, J. D., Henry, T. J., \& Irwin, M. J. 1997, \aj, 113, 1421

\bibitem[Kirkpatrick et al.(1991)]{kir91}
Kirkpatrick, J. D., Henry, T. J., \& McCarthy, D. W. 1991, \apjs, 77, 417

\bibitem[Lada \& Lada(1995)]{ll95}
Lada, E. A., \& Lada, C. J. 1995, \aj, 109, 1682

\bibitem[Landolt(1992)]{lan92}
Landolt, A. U. 1992, \aj, 104, 340

\bibitem[Leggett(1992)]{leg92}
Leggett, S. K. 1992, \apjs, 82, 351

\bibitem[Leggett et al.(2000)]{leg00}
Leggett, S. K., Allard, F., Dahn, C., Hauschildt, P. H., Kerr, T. H., 
\& Rayner, J. 2000, \apj, 535, 965

\bibitem[Leggett et al.(2001)]{leg01}
Leggett, S. K., Allard, F., Geballe, T. R., Hauschildt, P. H., \& Schweitzer,
A. 2001, \apj, 548, 908

\bibitem[Liu et al.(2003)]{liu03}
Liu, M. C., Najita, J., \& Tokunaga, A. T. 2003, \apj, 585, 372

\bibitem[Lucas et al.(2001)]{luc01}
Lucas, P. W., Roche, P. F., Allard, F., \& Hauschildt, P. H. 2001,
\mnras, 326, 695

\bibitem[Luhman(1999)]{luh99}
Luhman, K. L. 1999, \apj, 525, 466

\bibitem[Luhman(2000)]{luh00a}
Luhman, K. L. 2000, \apj, 544, 1044

\bibitem[Luhman(2001)]{luh01}
Luhman, K. L. 2001, \apj, 560, 287

\bibitem[Luhman et al.(1998a)]{luh98a}
Luhman, K. L., Brice\~{n}o, C., Rieke, G. H., \& Hartmann, L. W.
1998a, \apj, 493, 909

\bibitem[Luhman et al.(2003a)]{luh03}
Luhman, K. L., Brice\~{n}o, C., Stauffer, J. R., Hartmann, L., 
Barrado y Navascu\'{e}s, D., \& Nelson, C. 2003, \apj, in press

\bibitem[Luhman et al.(1997)]{luh97}
Luhman, K. L., Liebert, J., \& Rieke, G. H. 1997, \apj, 489, L165

\bibitem[Luhman et al.(2003b)]{lm03}
Luhman, K. L., McLeod, K. K., \& Goldenson, N. 2003, in preparation

\bibitem[Luhman \& Rieke(1998)]{lr98}
Luhman, K. L., \& Rieke, G. H. 1998, \apj, 497, 354

\bibitem[Luhman \& Rieke(1999)]{lr99}
Luhman, K. L., \& Rieke, G. H. 1999, \apj, 525, 440

\bibitem[Luhman et al.(1998b)]{luh98b}
Luhman, K. L., Rieke, G. H., Lada, C. J., \& Lada, E. A. 1998b, \apj, 508, 347

\bibitem[Luhman et al.(2000)]{luh00b}
Luhman, K. L., et al. 2000, \apj, 540, 1016

\bibitem[Mart{\'\i}n et al.(1996)]{mar96}
Mart{\'\i}n, E. L., Rebolo, R., \& Zapatero Osorio, M. R. 1996, \apj, 469, 706

\bibitem[McCaughrean et al.(1994)]{mcc94}
McCaughrean, M. J., Rayner, J. T., \& Zinnecker, H. 1994, \apj, 436, L189

\bibitem[Meyer et al.(1997)]{mey97}
Meyer, M. R., Calvet, N., \& Hillenbrand, L. A. 1997, \aj, 114, 288

\bibitem[Moraux et al.(2003)]{mor03}
Moraux, E., Bouvier, J., Stauffer, J. R., \& Cuillandre, J.-C. 2003, \aap, 400, 
891

\bibitem[Muench et al.(2002)]{mue02}
Muench, A. A., Lada, E. A., Lada, C. J., \& Alves, J. 2002, \apj, 573, 366

\bibitem[Muench et al.(2003)]{mue03}
Muench, A. A., et al. 2003, \aj, in press

\bibitem[Najita et al.(2000)]{naj00}
Najita, J., Tiede, G. P., \& Carr, J. S. 2000, \apj, 541, 977 (NTC00)

\bibitem[Oke et al.(1995)]{oke95}
Oke, J.B., et al. 1995, \pasp, 107, 375

\bibitem[Padoan \& Nordlund(2002)]{pn02}
Padoan, P., \& Nordlund, \AA. 2002, \apj, 576, 870

\bibitem[Palla \& Stahler(1999)]{pal99}
Palla, F., \& Stahler, S. W. 1999, \apj, 525, 772

\bibitem[Palla \& Stahler(2001)]{pal01}
Palla, F., \& Stahler, S. W. 2001, \apj, 553, 299

\bibitem[Popper(1987)]{pop87}
Popper, D. M. 1987, \apj, 313, 81

\bibitem[Prato(1998)]{pra98}
Prato, L. A. 1998, Ph. D. thesis, State Univ. of New York at Stony Brook

\bibitem[Prato et al.(2002)]{pra02}
Prato, L., Simon, M., Mazeh, T., McLean, I. S., Norman, D., \&
Zucker, S. 2002, \apj, 569, 863

\bibitem[Preibisch \& Zinnecker(2001)]{pre01}
Preibisch, T., \& Zinnecker, H. 2001, \aj, 122, 866

\bibitem[Preibisch \& Zinnecker(2002)]{pre02}
Preibisch, T., \& Zinnecker, H. 2002, \aj, 123, 1613

\bibitem[Preibisch et al.(1996)]{pre96}
Preibisch, T., Zinnecker, H., \& Herbig, G. H. 1996, \aap, 310, 456

\bibitem[Rieke \& Lebofsky(1985)]{rl85}
Rieke, G. H., \& Lebofsky, M. J. 1985, \apj, 288, 618

\bibitem[Ripepi et al.(2002)]{rip02}
Ripepi, V., Palla, F., Marconi, M., Bernabei, S., Arellano Ferro, 
A., Terranegra, L., \& Alcal\'a, J. M. 2002, \aap, 391, 587

\bibitem[Schmidt-Kaler(1982)]{sk82}
Schmidt-Kaler, T. 1982, in Landolt-Bornstein, Group VI, Vol. 2, ed. K.-H. 
Hellwege (Berlin: Springer), 454

\bibitem[Scholz et al.(1999)]{sch99}
Scholz, R.-D., et al. 1999, \aaps, 137, 305

\bibitem[Siess et al.(2000)]{sie00}
Siess, L., Dufour, E., \& Forestini, M. 2000, \aap, 358, 593

\bibitem[Simon et al.(2000)]{sim00}
Simon, M., Dutrey, A., \& Guilloteau, S. 2000, \apj, 545, 1034

\bibitem[Stahler(1983)]{sta83}
Stahler, S. W. 1983, \apj, 274, 822

\bibitem[Steffen et al.(2001)]{ste01}
Steffen, A. T., et al. 2001, \aj, 122, 997

\bibitem[Strom et al.(1974)]{str74}
Strom, S. E., Strom, K. M., \& Carrasco, L. 1974, \pasp, 86, 798

\bibitem[Tej et al.(2002)]{tej02}
Tej, A., Sahu, K. C., Chandrasekhar, T., \& Ashok, N. M. 2002, \apj, 578, 523

\bibitem[Torres \& Ribas(2002)]{tor02}
Torres, G., \& Ribas, I. 2002, \apj, 567, 1140

\bibitem[Walter et al.(1988)]{wal88}
Walter, F. W., Brown, A., Mathieu, R. D., Myers, P. C., \& Vrba,
F. J. 1988, \aj, 96, 297

\bibitem[White et al.(1999)]{whi99}
White, R. J., Ghez, A. M., Reid, I. N., \& Schultz, G. 1999, \apj, 520, 811

\bibitem[Williams et al.(1993)]{wil93}
Williams, D., Thompson, C. L., Rieke, G. H, \& Montgomery, E. F.
1993, ProcSPIE, 1308, 482

\end{thebibliography}
\end{document}